\documentclass[preprint,number]{elsarticle}

\usepackage{amsmath}
\usepackage{amssymb}
\usepackage{amsfonts}
\usepackage{amsthm}

\usepackage[normalem]{ulem}
\usepackage[T1]{fontenc}

\DeclareMathOperator{\Tr}{Tr}

\DeclareMathOperator{\hsv}{hsv}

\newtheorem{theorem}{Theorem}[section]
\newtheorem{corollary}[theorem]{Corollary}
\newtheorem{lemma}{Lemma}[section]
\newtheorem{definition}{Definition}[section]
\newtheorem{proposition}{Proposition}[section]
\newtheorem{example}{Example}[section]

\usepackage{graphicx}

\usepackage[caption=false]{subfig}
\usepackage{url}
\usepackage{color}

\biboptions{sort&compress,numbers}
\DeclareGraphicsExtensions{.eps,.jpg,.png}

\newcommand{\preprint}{
  \setlength{\unitlength}{1mm}{\hbox{\begin{picture}(0,0)
        \put(100,10){\mbox{\footnotesize%
            ADP-13-32/T852}}\end{picture}}}}

\begin{document}

\begin{frontmatter}

\title{\preprint
Visualisations of coherent center domains in local Polyakov loops}

\author[cssm]{Finn M. Stokes}
\ead{finn.stokes@adelaide.edu.au}
\author[cssm]{Waseem Kamleh}
\author[cssm]{Derek B. Leinweber}
\address[cssm]{Center for the Subatomic Structure of Matter,
School of Chemistry and Physics, University of Adelaide, SA 5005 Australia}
\date{\today}

\begin{abstract}
Quantum Chromodynamics exhibits a hadronic confined phase at
low to moderate temperatures and, at a critical temperature \(T_C\), undergoes a
transition to a deconfined phase known as the quark-gluon plasma.
The nature of this deconfinement phase transition is probed through
visualisations of the Polyakov loop, a gauge
independent order parameter. We produce visualisations that provide novel insights
into the structure and evolution of center clusters. Using the HMC algorithm the
percolation during the deconfinement transition is observed. Using 3D rendering of
the phase and magnitude of the Polyakov loop, the fractal structure and correlations
are examined. The evolution of the center clusters as the gauge fields thermalise
from below the critical temperature to above it are also exposed. We observe
deconfinement proceeding through a competition for the dominance of a particular
center phase. We use stout-link smearing to remove small-scale
noise in order to observe the large-scale evolution of the center clusters. A
correlation between the magnitude of the Polyakov loop and the proximity of its
phase to one of the center phases of SU(3) is evident in the visualisations.
\end{abstract}


\end{frontmatter}

\section{Introduction\label{sec:introduction}}
Quantum Chromodynamics (QCD) is the gauge field theory that describes the
strong interactions of quarks, the constituent
particles of hadrons such as protons and neutrons. In QCD, the strong
interaction is mediated by a gauge boson known as the gluon.
The self-coupling of gluons through the color charge gives rise to a
non-trivial vacuum structure, confining quarks and generating mass through
dynamical chiral symmetry breaking.

At a critical temperature, which in vacuum is \(T_C\approx 160\,\mathrm{MeV}\)
\cite{Bazavov:2011nk,Aoki:2006br,Aoki:2009sc,Borsanyi:2010bp} (around two
trillion degrees Kelvin), QCD undergoes a phase transition to a deconfined
phase. Above \(T_C\), confinement breaks down, resulting
in the formation of a quark-gluon plasma. Understanding the nature of this
transition is critical
to understanding the formation of hadronic matter in the early universe, the
nature of neutron stars, and observations at RHIC and the LHC
\cite{Asakawa:2012yv,Chatrchyan:2011sx,Aad:2010bu,Tonjes:2011zz}.

In this study, we work in pure \(SU(3)\) lattice gauge theory
in order to make quantitative comparisons with recent work
\cite{Gattringer:2010ug,Danzer:2010ge,Borsanyi:2010cw,Schadler:2013qba,Endrodi:2014yaa,Mykkanen:2012ri}.
In the absence of light dynamical quarks, the critical temperature increases to
\(T_C \approx 270 \, \mathrm{MeV}\) \cite{Karsch:1999vy}.

To observe this phase transition in Lattice QCD simulations, one examines the 
behavior of a complex-valued observable known as the Polyakov loop which acts
as an order parameter. It has an expectation value of zero in the confined phase
and a nonzero expectation value in the deconfined phase
\cite{Gattringer:2010ug}. As we will observe, this transition occurs through the
growth of center clusters, regions of space where the Polyakov loop
prefers a single complex phase
associated with the center of SU(3). It is these clusters that we analyse in
this paper.

We will demonstrate deconfinement in the behavior of the Polyakov loops at
\(T_C\) and produce and analyse visualisations of the center clusters
that allow the evolution of the center clusters to be directly observed for
the first time. We explore both the HMC evolution and the percolation as
\(T\) is increased above \(T_C\) and establish a correlation between the peaks
in the magnitude \(\rho(\vec{x})\) and the
proximity of \(\phi(\vec{x})\) to a center phase. This confirms the underlying
assumption of Ref.~\cite{Asakawa:2012yv}, which
links the center domain walls to phenomena observed at RHIC
and the LHC. This allows for a better understanding of the nature of the
phase transition in vacuum. To do this, we develop a custom volume renderer
that can correctly visualise the 3D complex field of local Polyakov loops.

\section{Background\label{sec:background}}
\subsection{Center Symmetry\label{sec:background:symmetry}}
In QCD, the gluons are described by the eight gluon fields \(A_\mu^a(x)\), which
are expressed as the sum
\[A_\mu(x) = A_\mu^a(x) \, T_a\,,\]
where \(T_a\) are the eight generators of SU(3). The local Polyakov loop is
defined to be
\begin{align*}
  L(\vec{x}) := \;&\textrm{Tr}\left(\mathcal{P}\exp \left[i g \int
                     \mathrm{d}x^0 A_0(x) \right]\right) \\
              = \;&\textrm{Tr}\prod_{t=1}^{N_t} U_0(t,\vec{x})\,,
\end{align*}
where \(U_0\) is the time-oriented link variable on a lattice with lattice
spacing \(a\), given by
\[U_{\mu}\left(x\right) = P\exp\left(ig\int_x^{x+\hat{\mu}a}dx^\mu A_\mu(x)\right)\,.\]
Under a local gauge transformation \(g(x) \in \text{SU(3})\),
\[U_{\mu}\left(x\right) \to g\left(x\right)U_{\mu}\left(x\right)
   g(x + \hat{\mu}a)^\dagger \, ,\]
so given periodic boundary conditions, the Polyakov loop is invariant under
local gauge transformations.

The spatially averaged Polyakov loop is
\[ P := \frac{1}{V} \sum_{\vec{x}}L(\vec{x})\,.\]

By translational invariance, the local Polyakov loop \(L(\vec{x})\) has the
same vacuum expectation value as the
spatially averaged loop \(P\). It is related to the free energy \(F_q\) of a
static quark by
\[\left<L(\vec{x})\right> = \left<P\right> \propto \exp(-F_q/T)\,.\]

In the confined phase, the free energy of a static quark is infinite, so
\(\left<L(\vec{x})\right> = \left<P\right> = 0\).
In the deconfined phase, the free energy of a static quark is finite, so
\(\left<L(\vec{x})\right> = \left<P\right> \ne 0\)
\cite{Danzer:2010ge}. Thus, the Polyakov loop is an order parameter for
confinement.

As it is defined here, the Polyakov loop would be exactly zero
in both phases, as in pure \(SU(N)\) gauge theory, all N sectors are equally
weighted in the partition function. Therefore usually the absolute value of
the Polyakov loop is used as an order parameter. An alternative approach,
which we take, is to remove this remaining symmetry by performing center
transformations to bring the most dominant center sector in each configuration
to the same phase. In full QCD the fermion determinant introduces a preferred
phase, causing the peak with a phase of zero to always become dominant above
the critical temperature \cite{Danzer:2010ge}, so this would no longer be a
concern.

In addition, on the lattice the expectation value of the absolute value of the
Polyakov loop below the critical temperature is not exactly zero, due to finite
volume effects. However, the volumes we work with are large enough that it is
close to zero. In the thermodynamic limit, the Polyakov loop truly vanishes below
the critical temperature.

If the center of the gauge group
\[C := \left\{z \in G | zg = gz \forall g \in G\right\}\]
is non-trivial, then the gauge action is invariant under a center transformation
\[U_0\left(t_0,\vec{x}\right) \to z U_0\left(t_0,\vec{x}\right) \forall \vec{x},
\text{ for some fixed } z \in C, t_0\,.\] 
These transformations form a global symmetry of the theory known as the center
symmetry. Polyakov loops transform non-trivially
under center transformations:
\[L(\vec{x}) \to z L(\vec{x})\,,\]
so if the center symmetry is conserved, \(\left<L(\vec{x})\right> =  0\)
\cite{Svetitsky:1982gs}. Thus, the Polyakov loop is an
order parameter for the center symmetry and deconfinement corresponds to a
spontaneous breaking of the center symmetry.

\begin{figure*}[t]
  \subfloat[\(T < T_C\), approximately equal occupation]{
    \label{fig:polyarg-cold}
    \includegraphics[width=0.47\textwidth]{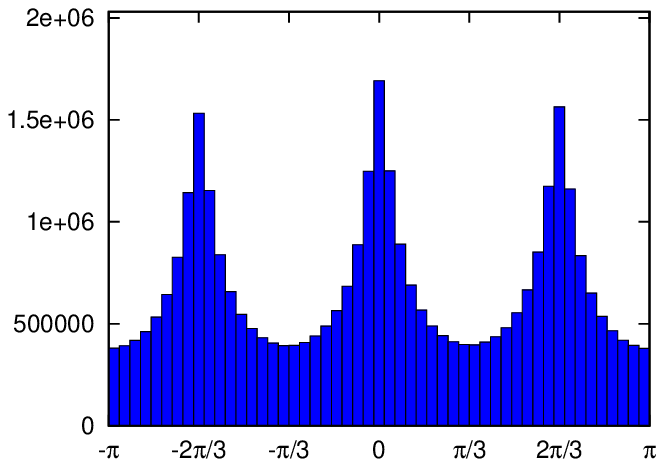}
  }
  ~
  \subfloat[\(T > T_C\), there is a single dominant center phase]{
    \label{fig:polyarg-hot}
    \includegraphics[width=0.47\textwidth]{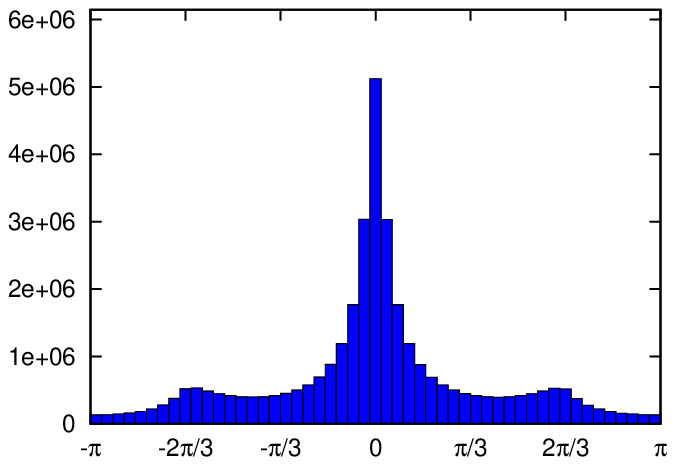}
  }
  \caption{Histograms of the distribution of complex phases of local loops
    across a gauge field ensemble, showing differing
    occupation of the three center phases.}
  \label{fig:polyarg}
\end{figure*}

It has been observed \cite{Danzer:2010ge,Gattringer:2010ug} that for SU(3) gauge
theory below the critical temperature, the complex
phase of the local Polyakov loops (\(\phi(\vec{x})\) in
\(L(\vec{x}) = \rho(\vec{x}) \, e^{i\phi(\vec{x})}\)) is distributed evenly
between three peaks, one at each of the three center phases of SU(3)
(\(0\), \(\frac{2\pi}{3}\), and \(\frac{-2\pi}{3}\)). We have
seen a similar effect in our own simulations as shown in
Fig.~\ref{fig:polyarg-cold}. As the distribution of \(L(\vec{x})\) is uniform
about the three center phases, the expectation value \(\left<L(\vec{x})\right>\)
vanishes. Above the
critical temperature, the center symmetry is spontaneously broken and one of the
three peaks becomes dominant.
We have replicated this in our simulations as shown in
Fig.~\ref{fig:polyarg-hot}. As a result, the expectation value
\(\left<L(\vec{x})\right>\) is non-zero.

\subsection{Anisotropic Gauge Action\label{sec:background:action}}
On the lattice, the temperature \(T\) is related to the temporal extent
\(a N_t\) by \(T = \frac{1}{a N_t}\).
The volume \(V = (a N_s)^3\) depends on the spatial extent of the lattice
\(a N_s\). Most studies of center domains change the temperature by varying the
lattice spacing \cite{Fortunato:2000fa,Gattringer:2010ug,Danzer:2010ge,
Borsanyi:2010cw}. This is a problem because the volume of the lattice is changed
as the temperature changes. The only way one can change the 
temperature of an isotropic lattice without changing the volume is by adding
or removing lattice sites in the time direction. In order to be able
to adjust the temperature
in a continuous manner, we instead introduce anisotropy into the lattice,
replacing the single lattice spacing \(a\) with a spatial
lattice spacing \(a_s\) and a temporal lattice spacing \(a_t\). In this way we
can vary the temperature \(T = \frac{1}{a_t N_t}\) while maintaining a constant
volume \(V = (a_s N_s)^3\) by varying \(a_t\) whilst holding \(a_s\) fixed. This
allows us to observe the evolution of the center clusters in gauge field
configurations moving from a confined configuration to a deconfined
configuration under the HMC algorithm with no change in physical volume. This
cannot be done with any of the previous methods and is the first such
presentation.

In order to introduce this anisotropy, we use the anisotropic Iwasaki
action~\cite{Umeda:2003pj}, which introduces an anisotropy parameter
\(\gamma_G\). This separates the Wilson loops in the plane of
two spatial directions from the Wilson loops in the plane of a single spatial
and temporal direction. The gauge action is
\begin{align*}
  S_G[U] &= \beta \Bigg\{\frac{1}{\gamma_G} \sum_{x,i>j} \left\{c_0^s P_{ij}(x)
          + c_1^s (R_{ij}(x) + R_{ji}(x))\right\} \\
         &+ \gamma_G \sum_{x,k} \left\{c_0^t P_{k4}(x) + c_1^t R_{k4}(x)
          + c_2^t R_{4k}(x)\right\}\Bigg\}\,,
\end{align*}
where \(i\), \(j\), \(k\) are for spatial directions, and \(P_{\mu\nu}(x)\) and
\(R_{\mu\nu}(x)\) are the \(1\times1\) plaquette
and \(1\times2\) rectangular loop in the \(\mu-\nu\) plane respectively,
\(\beta\) governs the strong coupling, \(\gamma_G\) governs
the anisotropy, and the improvement coefficients are \cite{Umeda:2003pj}
\begin{align*}
  c_1^s = c_1^t = c_2^t &= -0.331 \\
  c_0^s = c_0^t &= 3.648\,.
\end{align*}

\subsection{Hybrid Monte-Carlo\label{sec:background:hmc}}
We want to generate an ensemble of gauge field configurations
\(\left\{U_i\right\}\) with probability distribution
\(\rho(U) = e^{-S_G[U]}\). To do this, it is common to use local
updates such as the pseudo heat-bath combined with over-relaxation algorithms as
this is an efficient way to generate pure gauge field configurations. However, we
are ultimately interested in dynamical fermion results, so we use the Hybrid
Monte Carlo (HMC) algorithm \cite{Duane1987216}. This allows us to study the way
the HMC algorithm evolves the gauge fields at fixed temperature and from a state
thermalised at one temperature to a new state eventually thermalised at a higher
temperature. This is of interest to the lattice community who use the HMC
algorithm for generating dynamical gauge field ensembles, as it gives insight
into the nature of the algorithm and correlation times.

The HMC algorithm involves introducing the non-physical constructs \(\Pi_\mu\)
and \(\tau\), the conjugate momenta of \(U\) and the
simulation time respectively. We then describe the new extended system by the
Hamiltonian
\[\mathcal{H}[U,\Pi] = \sum_{x,\mu} \frac{1}{2}\, \Tr \Pi_\mu(x)^2 + S_G[U]\, .\]

To describe the frame rate of our visualisations, we briefly review the update
algorithm.
Starting from an initial gauge field configuration \(U\), we apply the following
process \cite{Kamleh:2004xk}:

\begin{enumerate}
  \item Select a random \(\Pi\) from an ensemble distributed according to
    \(\rho(\Pi) \propto e^{-\frac{1}{2} \mathrm{Tr} \Pi^2}\)
  \item Perform molecular dynamics updates with step size \(\Delta \tau\),
    evolving \(U_\mu\) to to \(U_\mu'\) and
    \(\Pi_\mu\) to \(\Pi_\mu'\) by the following discretised equations of motion 
    (so that
    \(\frac{d\mathcal{H}}{d\tau} \approx 0\)):
    \[U_\mu(x,\tau+\Delta\tau) = U_\mu(x,\tau)\exp(i\Delta\tau \Pi_\mu(x,\tau))\]
    \[\Pi_\mu(x,\tau+\Delta\tau) = \Pi_\mu(x,\tau)
        - \Delta\tau U_\mu(x,\tau)\frac{\delta S_G}{\delta U_\mu(x,\tau)}\]
  \item After \(\frac{1}{\Delta\tau}\) updates, providing a trajectory length of
    1, we accept or reject the new
    configuration with probability \(\rho = \min(1,e^{-\Delta\mathcal{H}})\), where
    \(\Delta\mathcal{H} = \mathcal{H}[U,\Pi] - \mathcal{H}[U',\Pi']\).
\end{enumerate}

In generating the configurations used in this paper, we used 150 molecular
dynamics steps with step size
\(\Delta \tau = \frac{1}{150}\). When producing
visualisations we store the state of the Polyakov loops five times per
trajectory in order to make the animation
smoother. Of course, these frames are dropped if the configuration is not
accepted in step 3. The acceptance rates we observed when generating our
gauge field configurations varied between 75\% and 90\%, depending on the
volume and temperature of the configurations being generated. Large volumes
and high temperatures have the lowest acceptance rates.

Independent gauge fields are saved every 50 trajectories, an
order of magnitude more separation than the 5 trajectories often used in
dynamical QCD.

As we will see, correlations in the centre clusters are associated with a time
scale of 20 to 30 trajectories, or 4 to 6 seconds in the animations.

\subsection{Lattice Spacings and Temperature\label{sec:background:spacing}}
The lattice spacings \(a_s\) and \(a_t\) depend non-trivially on \(\beta\) and
\(\gamma_G\). In order to determine \(a_s\) and \(a_t\)
for a given \((\beta, \gamma_G)\) pair, we perform a lattice simulation at zero
temperature (high temporal extent) and fit the
values of the Wilson loops, \(W_{\mu\nu}(r,t)\), to the static quark potential
\[V(r) = V_0 + \sigma r - e \cdot \left[\frac{1}{r}\right]
         + l \cdot \left(\left[\frac{1}{r}\right]-\frac{1}{r}\right) \,,\]
where \(\left[\frac{1}{r}\right]\) denotes the tree-level lattice Coulomb term
\cite{Edwards:1997xf}.

This ansatz for the static quark potential is sensitive to
discretisation effects at extremely small \(r\) and noise starts to dominate at
large \(r\), so we need a
lower and upper cutoff for \(r\).

On an anisotropic lattice, the fit for the space-space loops
(\(W_{xy}(r,t)\), \(W_{xz}(r,t)\), and \(W_{yz}(r,t)\))
gives a string tension \(\sigma_{ss}\) which is related to the spatial lattice
spacing \(a_s\) through the physical value
\(\sqrt{\sigma} = 0.44\,\mathrm{GeV}\).
On the other hand, the space-time loops,
(\(W_{xt}(r,t)\), \(W_{yt}(r,t)\), and \(W_{zt}(r,t)\))
and the time-space loops (\(W_{tx}(r,t)\),
\(W_{ty}(r,t)\), and \(W_{tz}(r,t)\)) give \(\sigma_{st}\), which is related to
both the spatial and temporal lattice spacings via
\(\sqrt{a_sa_t}\).

\begin{table}[tb]
\caption{The \((\beta, \gamma_G)\) pairs obtained from pure gauge simulations
  of \(24^3 \times 48\) lattices and the corresponding temperatures for a
  \(24^3 \times 8\) lattice.
  \label{tab:temp}}
\begin{tabular}{cccccc}
\(\beta\) & \(\gamma_G\) & \(a_s\) (fm) & \(a_t\) (fm) & \(\xi\) & \(T\) (MeV)\\
\hline
2.620 & 1.000 & 0.1016(5)  & 0.1028(11) & 0.988(12) & 240(3) \\
2.645 & 1.125 & 0.1014(4)  & 0.0912(8)  & 1.112(12) & 270(2) \\
2.670 & 1.250 & 0.1019(6)  & 0.0803(10) & 1.245(19) & 307(4) \\
2.695 & 1.375 & 0.1014(4)  & 0.0761(8)  & 1.335(25) & 324(3) \\
2.720 & 1.500 & 0.1002(9)  & 0.0673(10) & 1.489(31) & 366(5) \\
2.740 & 1.625 & 0.1007(12) & 0.0622(8)  & 1.618(39) & 396(5) \\
2.760 & 1.750 & 0.1019(8)  & 0.0567(9)  & 1.796(38) & 434(7) \\
\end{tabular}
\end{table}

By finding \(a_s\) and \(a_t\) for a variety of \((\beta, \gamma_G)\) pairs,
the relationship between
\(\beta\) and \(\gamma_G\) necessary to keep \(a_s\) constant can be determined.
This allows us to choose a set of
\((\beta, \gamma_G)\) pairs that give us access to a range of temperatures at
a fixed volume as shown in Table~\ref{tab:temp}.

We also found that the renormalised anisotropy,
\[\xi = \frac{a_s}{a_t} \, ,\]
is approximately equal to the bare anisotropy, \(\gamma_G\), in the region we
studied.

\subsection{Potts Model}
The universal properties at finite temperature phase transitions in (3 + 1)
dimensional gauge theories are related to those in 3-dimensional spin
models \cite{Karsch:2001ya,Svetitsky:1982gs}. Thus, it is of interest to compare
the behavior of the Polyakov loops in QCD with the three
dimensional 3-state Potts model \cite{Janke:1996qb}, a generalisation of the
Ising spin model. In the 3-state Potts model, each lattice
site can assume one of three spin directions and these form spin aligned
domains. After the phase transition, one direction
dominates the space, just as in the QCD deconfinement transition. Thus, the
Potts model is a candidate for a simplified model of
deconfinement.

The 3-state Potts model for a three dimensional lattice of points \(\vec{x} \in 
\mathbb{L} \subset \mathbb{Z}^3\) with spin \(\sigma(\vec{x}) =
1,2\text{ or }3\)
at each lattice site is described by the partition function
\[Z = e^{-\beta E[\sigma]}\, ,\]
where
\[E[\sigma] = \sum_{\vec{x}} \sum_i \delta_{\sigma(\vec{x}),\sigma(\vec{x}+\hat{\imath})}\, ,\]
is summed over the points of the lattice and the three spatial directions,
\(\hat{\imath}\) is the unit vector in the positive \(i\) direction, and
\(\delta_{j,k}\) is the Kronecker delta. That is, \(E[\sigma]\) is the total
number of spin-aligned nearest-neighbor pairs.
Here, \(\beta = J/k_BT\) is the inverse temperature in natural units.

We can simulate this using the Metropolis-Hastings algorithm
\cite{Hastings197097} to study the behavior of spin-aligned domains near the
critical temperature and compare it to the behavior of center clusters in QCD
near \(T_C\).

\section{Results\label{sec:results}}
\subsection{Visualisation\label{sec:results:visualisation}}
In order to analyse the behavior of the Polyakov loops at the critical
temperature, we generate gauge field
configurations on several \(24^3 \times 8\) lattices with the same spatial
lattice spacing but different temporal spacings,
using the parameters described in Section~\ref{sec:background}.

\begin{figure*}[tb]
  \subfloat[Interpolation of phase]{
    \label{fig:rigraph}
    \includegraphics[width=0.47\textwidth]{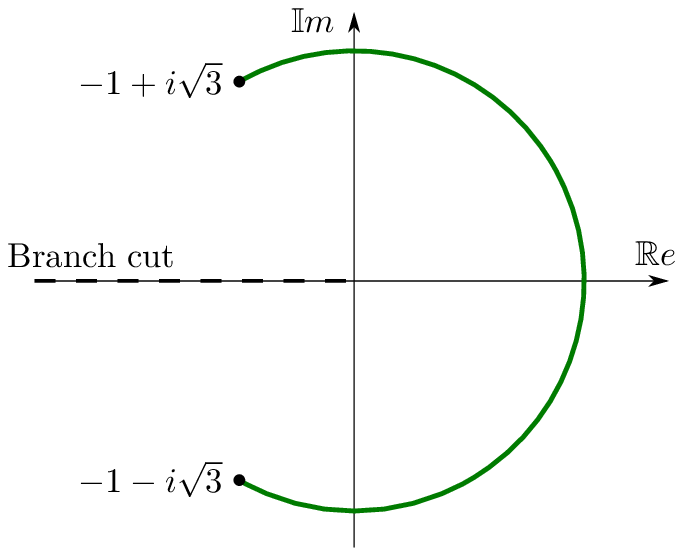}
  }
  \qquad
  \subfloat[Interpolation of complex value]{
    \label{fig:cigraph}
    \includegraphics[width=0.47\textwidth]{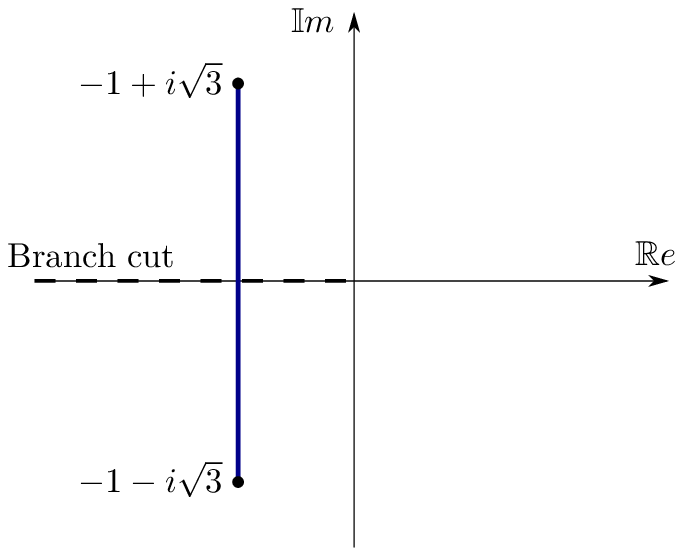}
  }
  \caption{Diagrams showing path followed by different interpolation methods.}
  \label{fig:interpolation}
\end{figure*}

\begin{figure*}[tb]
  \subfloat[Interpolation of phase (artifacts)]{
    \label{fig:rirender}
    \includegraphics[width=0.47\textwidth]{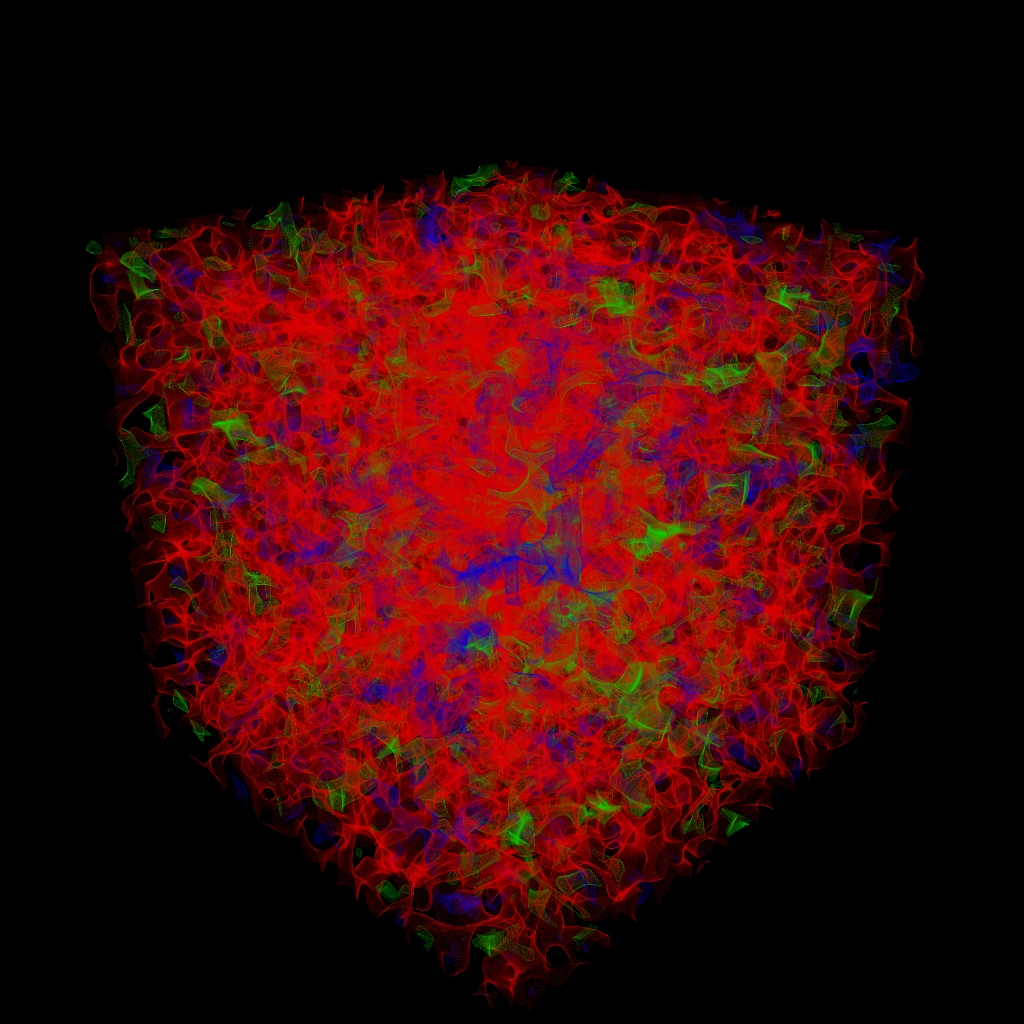}
  }
  \qquad
  \subfloat[Interpolation of complex value]{
    \label{fig:cirender}
    \includegraphics[width=0.47\textwidth]{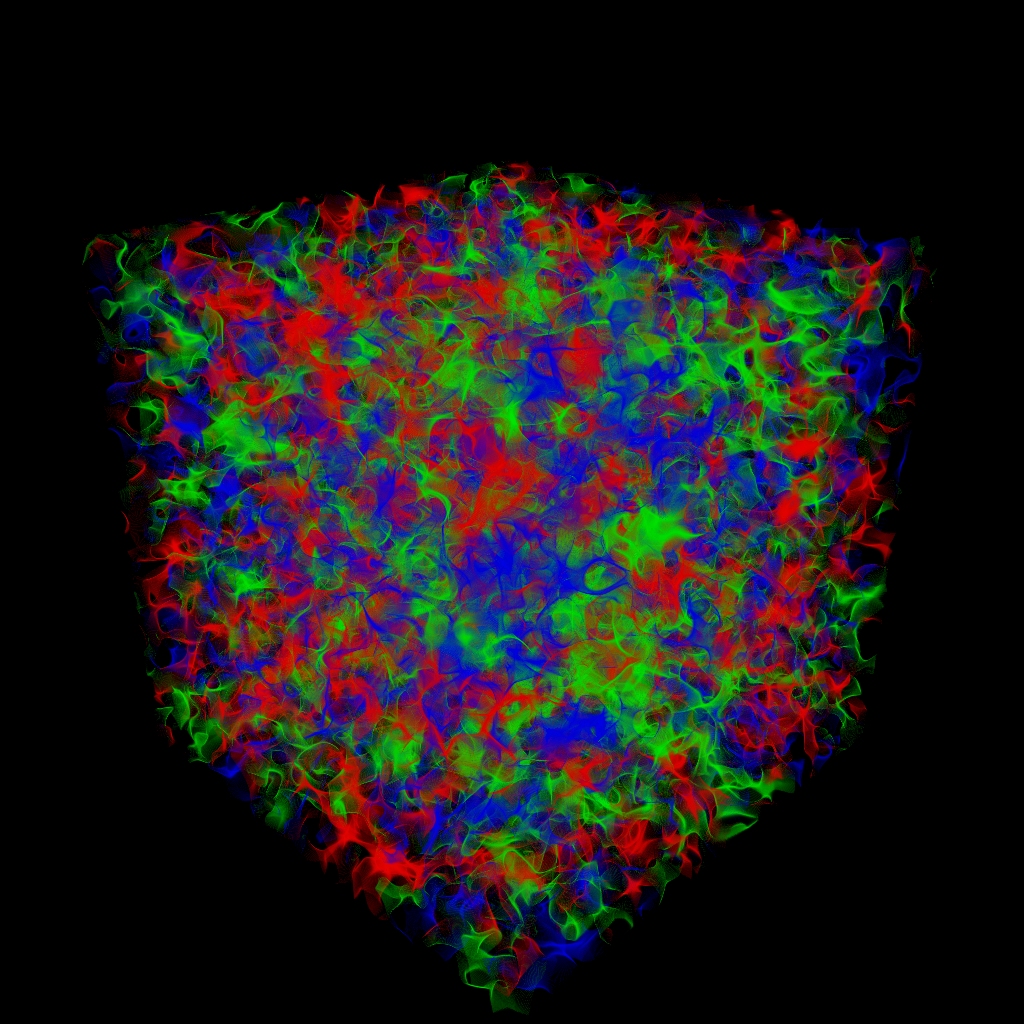}
  }
  \caption{Visualisations demonstrating the (red) artifacts
    produced by the
    incorrect phase interpolation on the left. On the right, the correct
    complex interpolation shows an equal center phase distribution.}
  \label{fig:render}
\end{figure*}

Visualising the Polyakov loop data presents a difficulty. Most visualisation
software that
has the ability to render volumetric data takes in a three dimensional grid of
real values and uses trilinear
interpolation to fill in the gaps. However, if we try to do this with the
complex phase of our Polyakov
loop values (with a branch cut at \(\pi\)), we find that the interpolation has
problems
near the branch cut. For example, if two adjacent data points have complex
values
\(-1 \pm i\sqrt{3}\) (corresponding to complex phases of
\(\phi = \frac{\pm 2\pi}{3}\)) then interpolating the phase linearly
between the two points takes it through \(\phi = 0\) as in
Fig.~\ref{fig:rigraph}, rather than crossing the branch cut as in
Fig.~\ref{fig:cigraph}. This behavior is incorrect
and produces artifacts in the final image that look like thin shells or films
between regions of transparency.
This leads to much more red (corresponding to \(\phi = 0\)) than either of green
or blue (\(\phi = \frac{2\pi}{3}\) and
\(\phi = \frac{- 2\pi}{3}\) respectively) as seen in Fig.~\ref{fig:rirender}. 

Instead, we want to directly interpolate the complex numbers and
\emph{then} calculate the
phase, resulting in the phase going directly between the two endpoints, across
the branch cut as in
Fig.~\ref{fig:cigraph}. This results in a significantly different visualisation
as seen in
Fig.~\ref{fig:cirender}.

To circumvent this problem, we have developed a custom volume renderer for
complex valued scalar fields. The details of the rendering algorithm are given
in \ref{sec:visualisation}. There we present two ways of visualising
the center clusters,
one based on proximity of \(\phi(\vec{x})\) to one of the three center phases
and one on the magnitude \(\rho(\vec{x})\).

\subsection{Critical Temperature\label{sec:results:crittemp}}
Four animations are available online at the URL provided in
Ref.~\cite{animations}. This is the first presentation of animations of the
evolution of center clusters under the HMC algorithm. Two different rendering
techniques are presented. As discussed in \ref{sec:visualisation}, these
include rendering based on the proximity of
\(\phi(\vec{x})\) to one of the
three center phases of SU(3) and rendering based on the magnitude
\(\rho(\vec{x})\). Each are rendered from the original configurations
as well as configurations smoothed with four sweeps of stout-link smearing as
discussed in greater detail below.

\begin{figure*}[ptb]
  \subfloat{
    \includegraphics[width=0.57\textwidth]{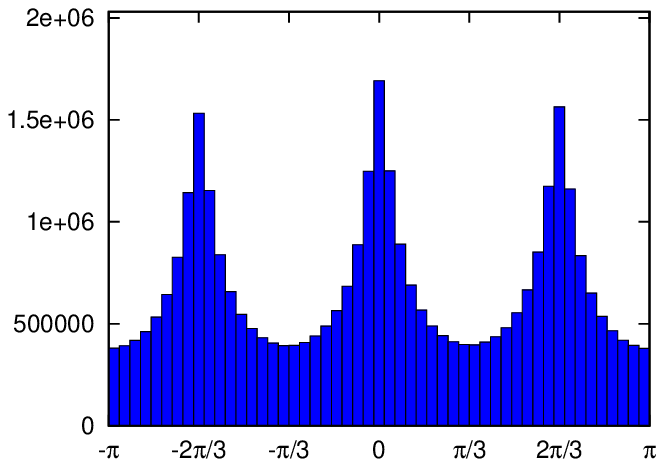}
    \label{fig:hist1.0}
  }
  \qquad
  \subfloat{
    \includegraphics[width=0.37\textwidth]{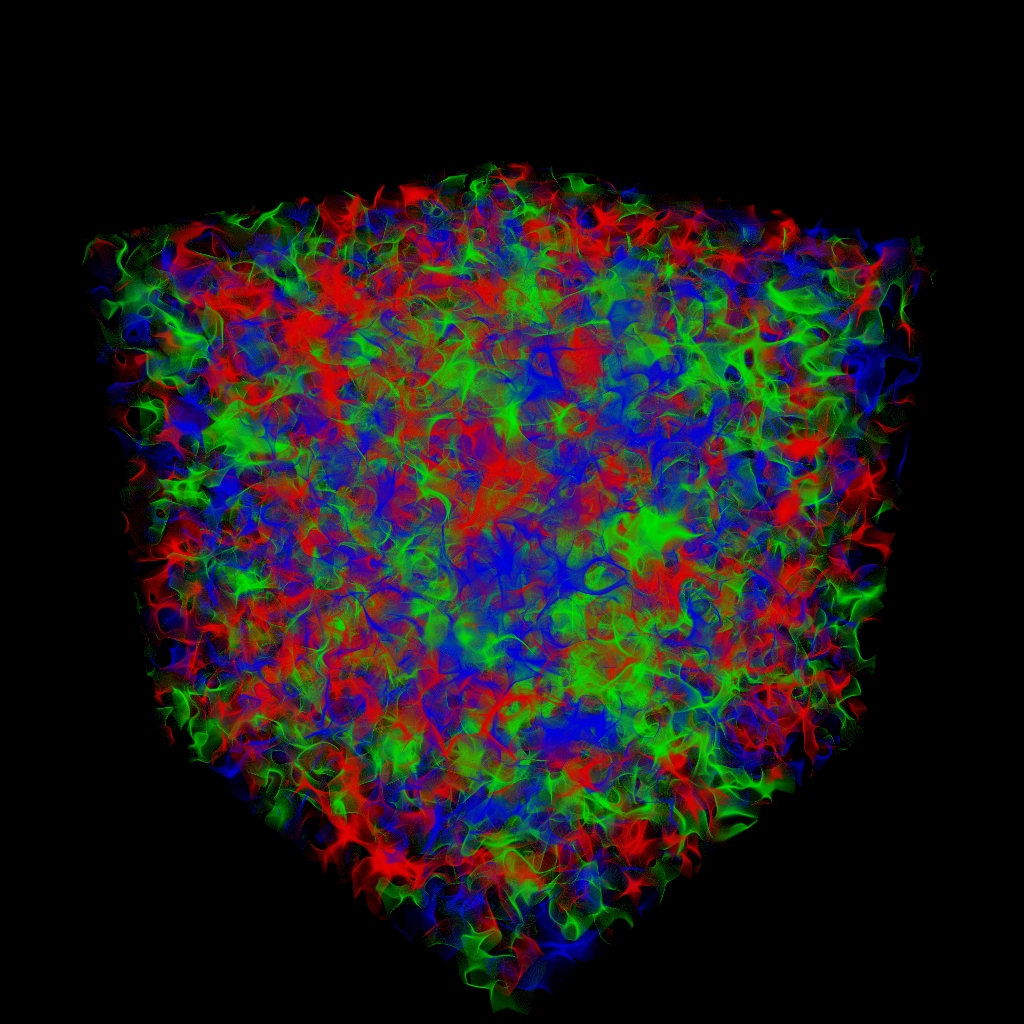}
    \label{fig:vis1.0}
  }
  \caption{Histogram of the phase \(\phi(\vec{x})\) and
    a visualisation of the
    center clusters in the local Polyakov loop values at
    \(T = 0.89(1) \, T_C\). All three center phases are present in small
    clusters with approximately equal density.}
  \label{fig:xi1.0}
\end{figure*}

The animations reveal the evolution of center clusters as a function of the
HMC simulation time, with five frames per unit trajectory. The
evolution of the center clusters is governed by the evolution of the gauge
field configurations under the HMC algorithm, and therefore the timescale of
these animations is not a physical scale but an algorithmic scale. However,
while it may have no direct physical significance, the evolution of the center
clusters during thermalisation gives novel insights into the nature of the HMC
algorithm and the way it brings the gauge field to a physical configuration. In
addition, once thermalisation is complete, every frame in itself is a physically
representative state and thus we can observe a number of possible structures for
the center clusters and gain extensive insight.

We commence with a thermalised configuration at \(T = 240(3) \,\mathrm{MeV}\) or
\(T/T_C = 0.89(1)\). We select a \(24^3\) spatial volume and note that it is
representative of the other volumes considered herein.
To show the nature of the HMC updates, we
present 750 frames corresponding to 150 HMC trajectories. At our framerate of 25
frames per second, this lasts 30 seconds. A snapshot of
the center clusters at this temperature is provided in Fig.~\ref{fig:xi1.0}. On
the left-hand plot, a histogram of the distribution of the phases of the
Polyakov loops across a gauge field ensemble at this temperature shows that all
three center phases, observable as peaks in the histogram, are approximately
equally occupied, signaling confinement.
The right-hand plot shows a single frame from the animation, in which the three
center-phase peaks observed in the histogram correspond to blue (left),
red (center), and green (right). All three center phases are present in small
clusters with approximately equal density.

\begin{figure*}[ptb]
  \subfloat{
    \includegraphics[width=0.57\textwidth]{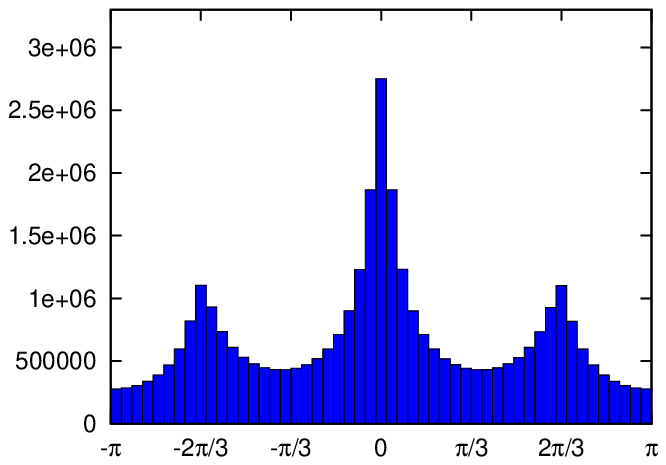}
    \label{fig:hist1.25}
  }
  \qquad
  \subfloat{
    \includegraphics[width=0.37\textwidth]{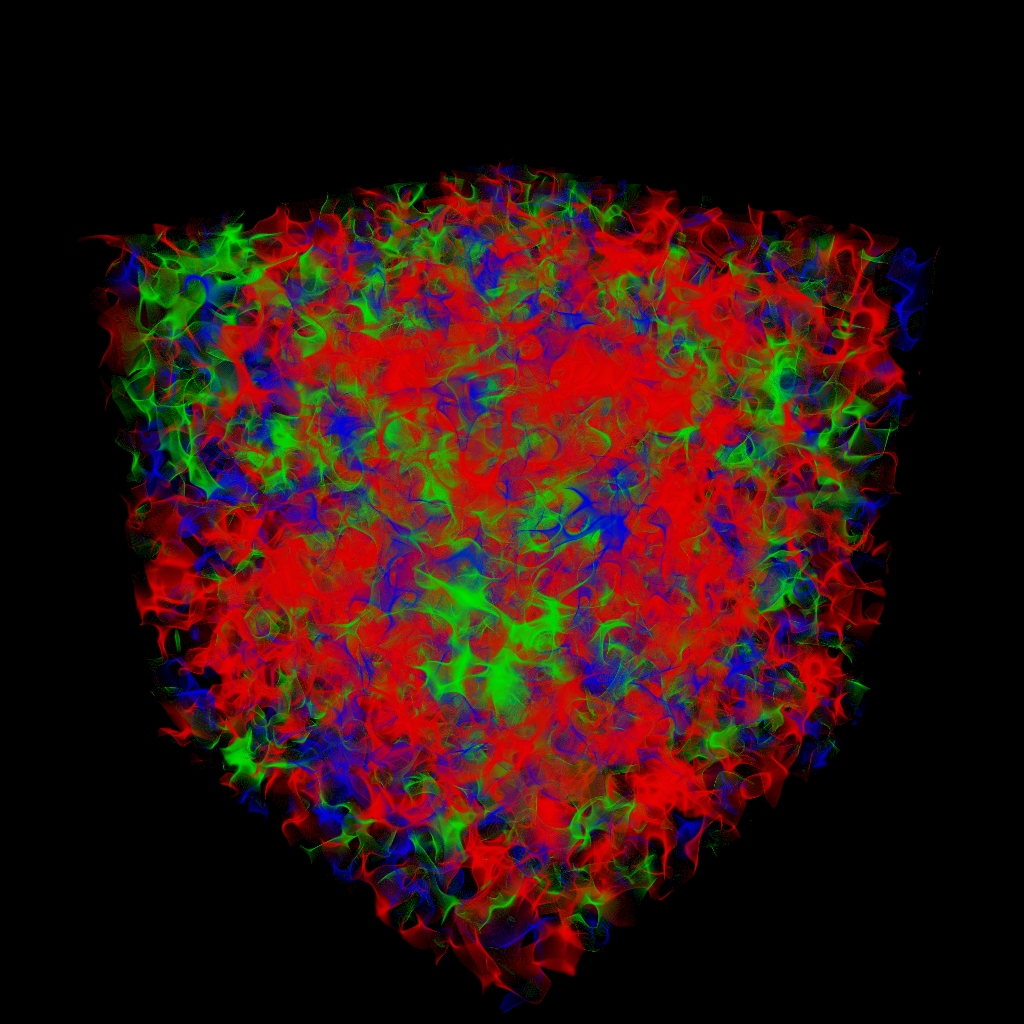}
    \label{fig:vis1.25}
  }
  \caption{Histogram and visualisation as in
    Fig.~\ref{fig:xi1.0} at
    \(T = 1.14(2) \, T_C\). A single(red) phase is beginning to dominate,
    signaling deconfinement.}
  \label{fig:xi1.25}
\end{figure*}

At this point (0:30 in the animation), the temperature is increased to
\(T = 307(4) \,\mathrm{MeV}\) or \(T/T_C = 1.14(2)\) and the response of
the gauge field is illustrated
in the animations. Fig.~\ref{fig:xi1.25} shows the red center phase becoming
dominant with the other two beginning to be suppressed, signaling
the onset of deconfinement. In the animation, we can see that the three
center phases start out equally dominant and fluctuate in size until the
red clusters come to dominate.

\begin{figure*}[ptb]
  \subfloat{
    \includegraphics[width=0.57\textwidth]{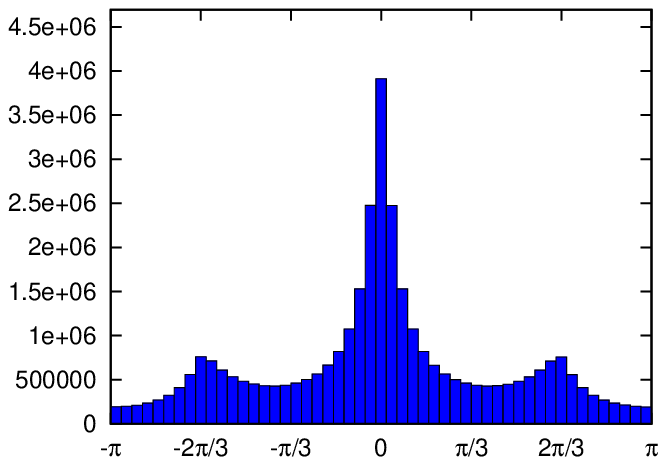}
    \label{fig:hist1.5}
  }
  \qquad
  \subfloat{
    \includegraphics[width=0.37\textwidth]{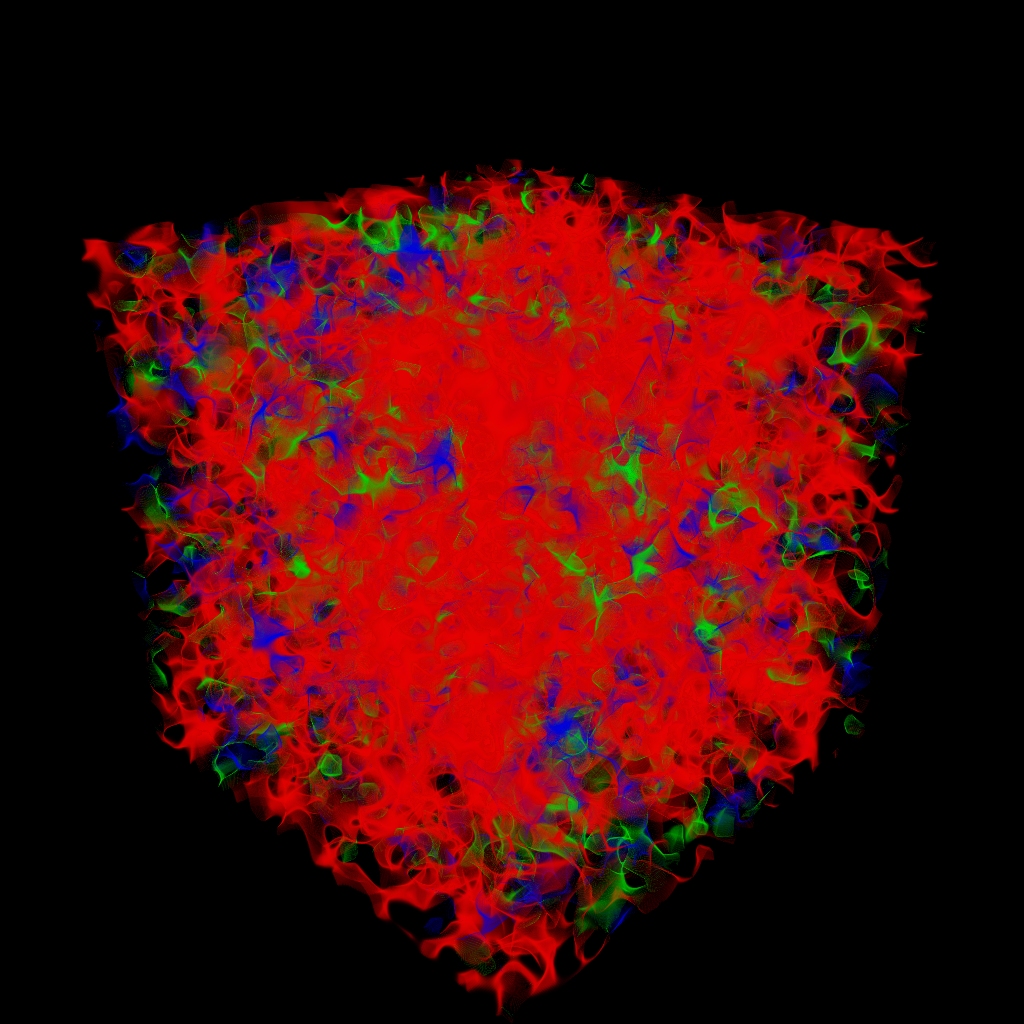}
    \label{fig:vis1.5}
  }
  \caption{Histogram and visualisation at
    \(T = 1.36(2) \, T_C\). A single (red) cluster dominates the entire space.}
  \label{fig:xi1.5}
\end{figure*}

After 600 HMC trajectories or 120 seconds, at 2:30 in the animation, the
configuration has responded to the temperature change and we increase the
temperature again, this
time to \(T = 366(5) \,\mathrm{MeV}\) or \(T/T_C = 1.36(2)\). At this
temperature, the animation shows that the red clusters continue to
grow, occupying almost
the entire space. This can be seen in the snapshot provided in
Fig.~\ref{fig:xi1.5}.

\begin{figure*}[ptb]
  \subfloat{
    \includegraphics[width=0.57\textwidth]{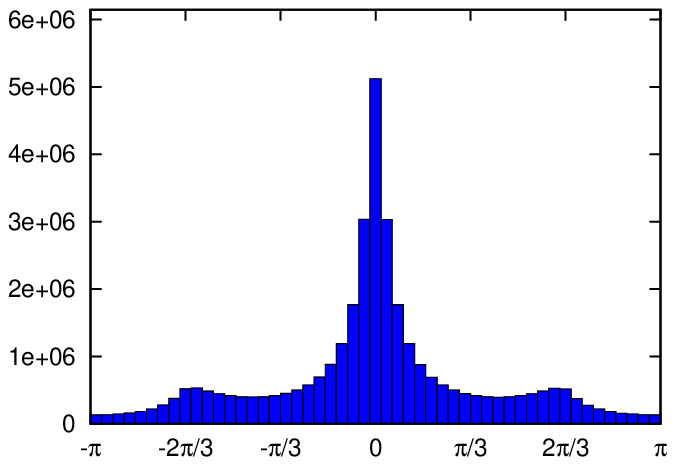}
    \label{fig:hist1.75}
  }
  \qquad
  \subfloat{
    \includegraphics[width=0.37\textwidth]{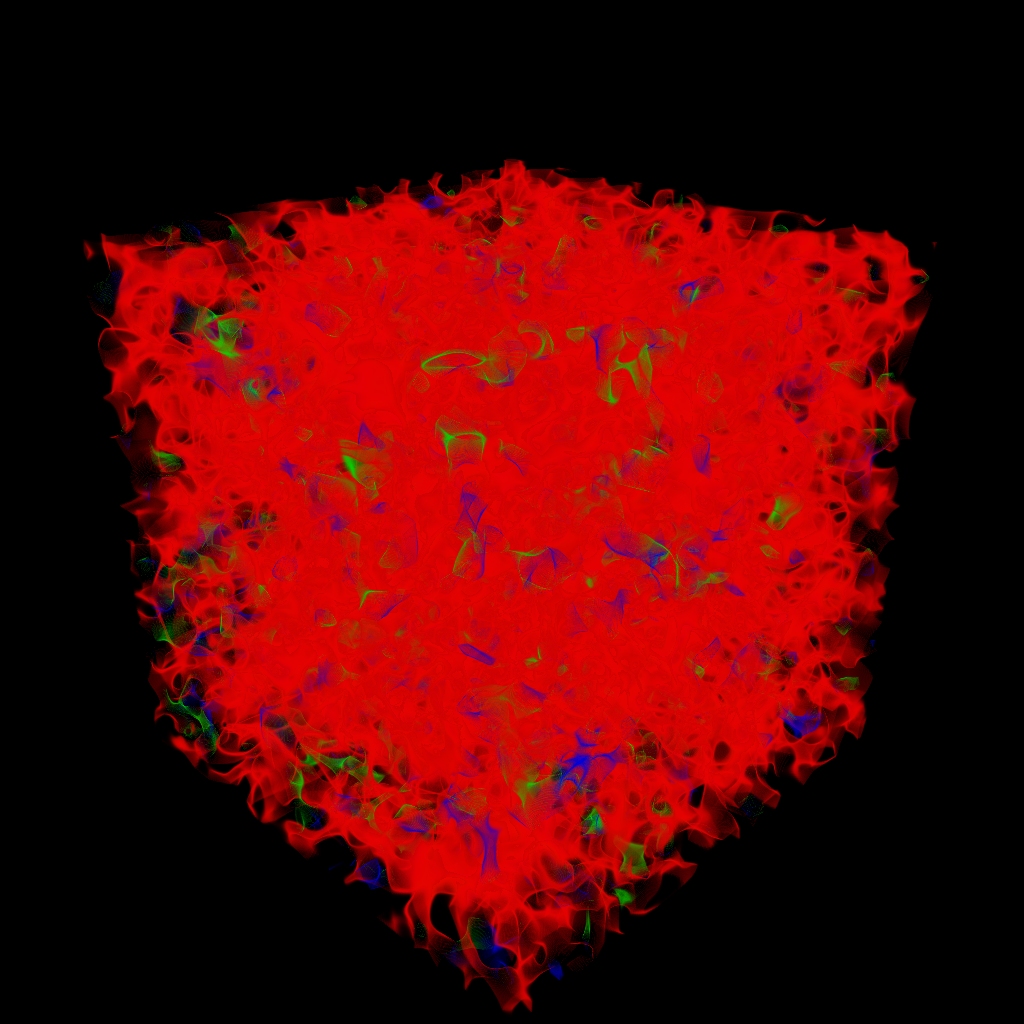}
    \label{fig:vis1.75}
  }
  \caption{Histogram and visualisation at
    \(T = 1.61(3) \, T_C\). The same (red) phase is still dominant.}
  \label{fig:xi1.75}
\end{figure*}

After 250 more HMC trajectories or 50 seconds, at 3:20 in the animation, we
increase the temperature again, to \(T = 434(7) \,\mathrm{MeV}\) or
\(T/T_C = 1.61(3)\). At this temperature,
the red phase remains dominant in the animation and the other two phases are
suppressed even further. Fig.~\ref{fig:xi1.75} provides a snapshot at this
temperature, showing the red phase almost completely dominant. We show 250 HMC
trajectories at this temperature over the remaining 50 seconds of the animation.

\subsection{Monte-Carlo Evolution\label{sec:results:evol}}
\begin{figure*}[ptb]
  \subfloat[Thermalised configuration]{
    \includegraphics[width=0.47\textwidth]{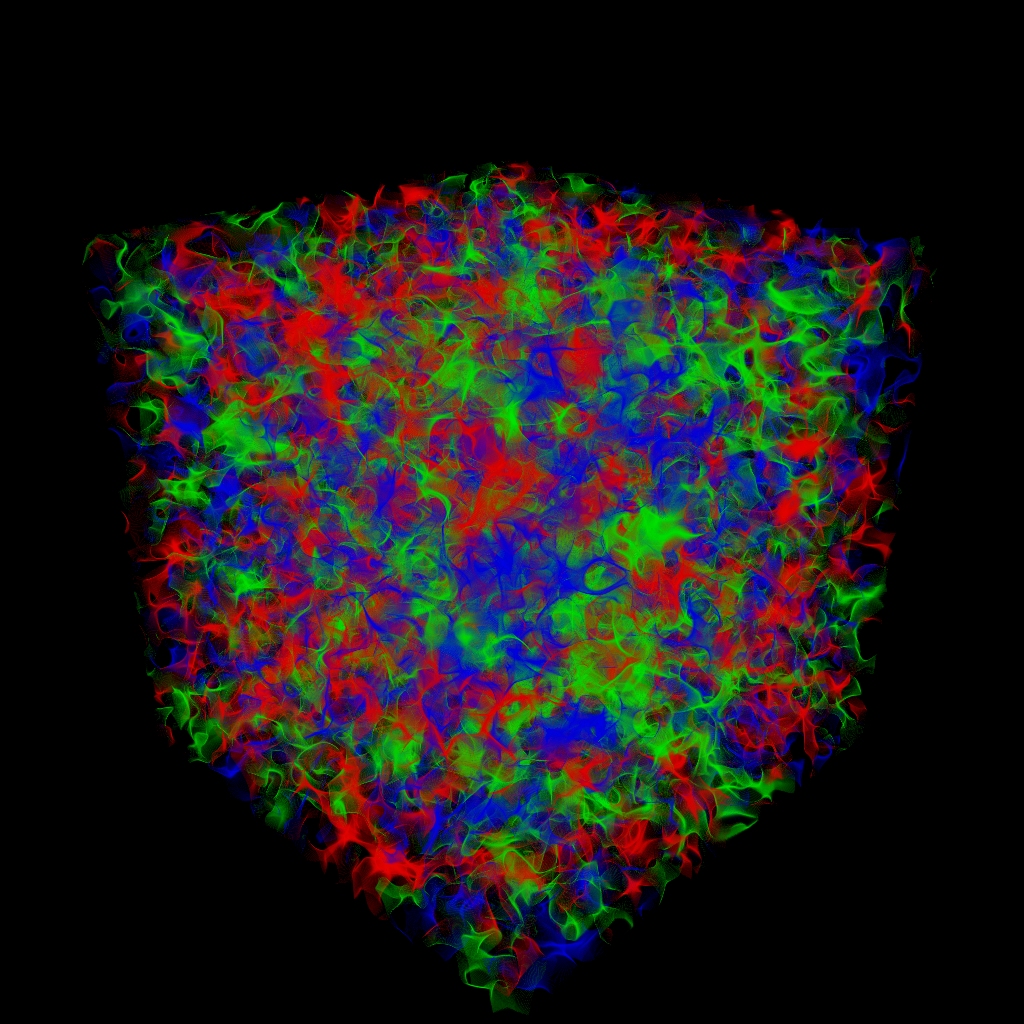}
    \label{fig:evolution1}
  }
  \qquad
  \subfloat[After one unit HMC trajectory]{
    \includegraphics[width=0.47\textwidth]{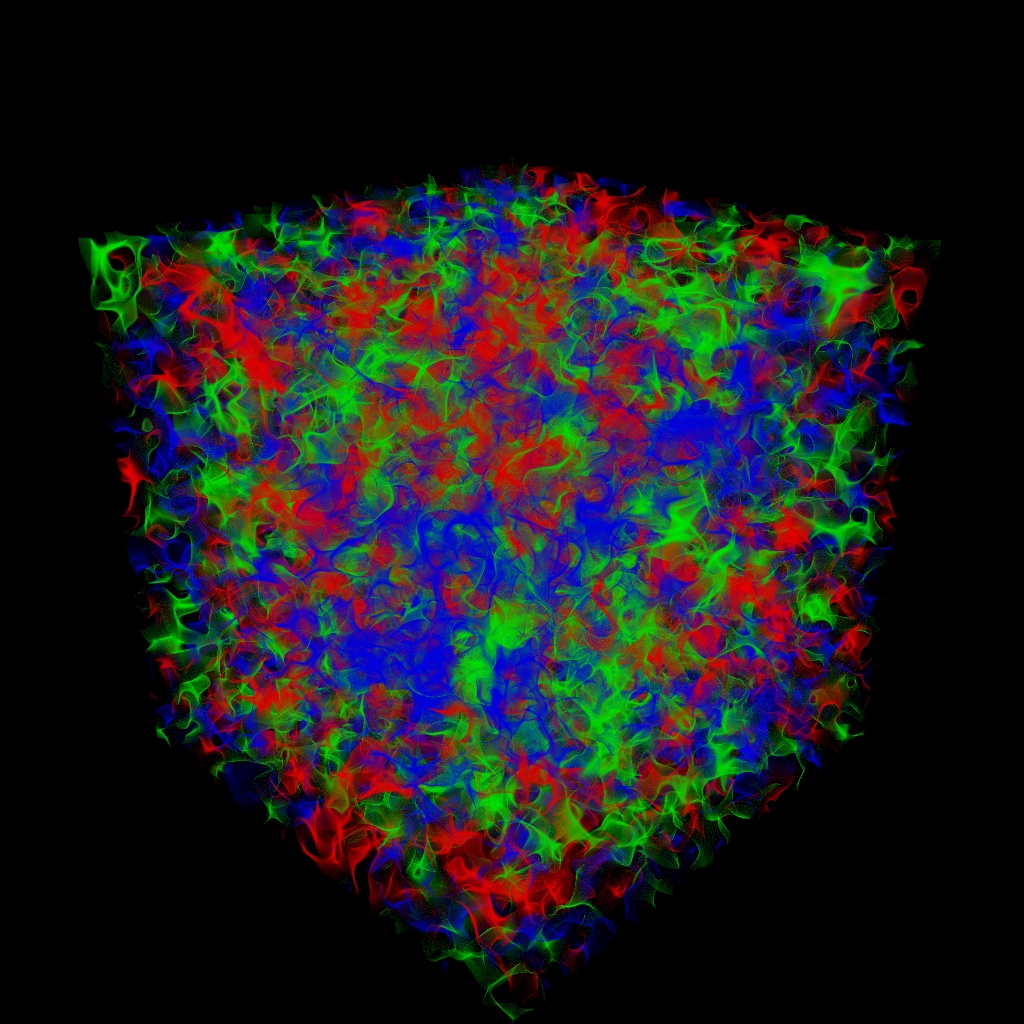}
    \label{fig:evolution2}
  }
  \caption{Evolution of center clusters with simulation time at
    \(T = 0.89(1) \, T_C\).}
  \label{fig:evolution}
\end{figure*}

By examining visualisations of the center clusters on gauge field configurations
separated by a single HMC trajectory, we can
observe the evolution of the center clusters with simulation time. The
autocorrelation times for different observables under HMC evolution can vary,
so it is of interest to observe the
timescales over which the center clusters evolve.
We can see that over the course of a single unit trajectory, the small scale
structure of the loops changes significantly, as shown in
Fig.~\ref{fig:evolution}. This suggests that the small scale structure
of the center clusters evolves very quickly under the HMC algorithm and thus has short
autocorrelation times.

\begin{figure*}[ptb]
  \subfloat[Thermalised configuration]{
    \includegraphics[width=0.47\textwidth]{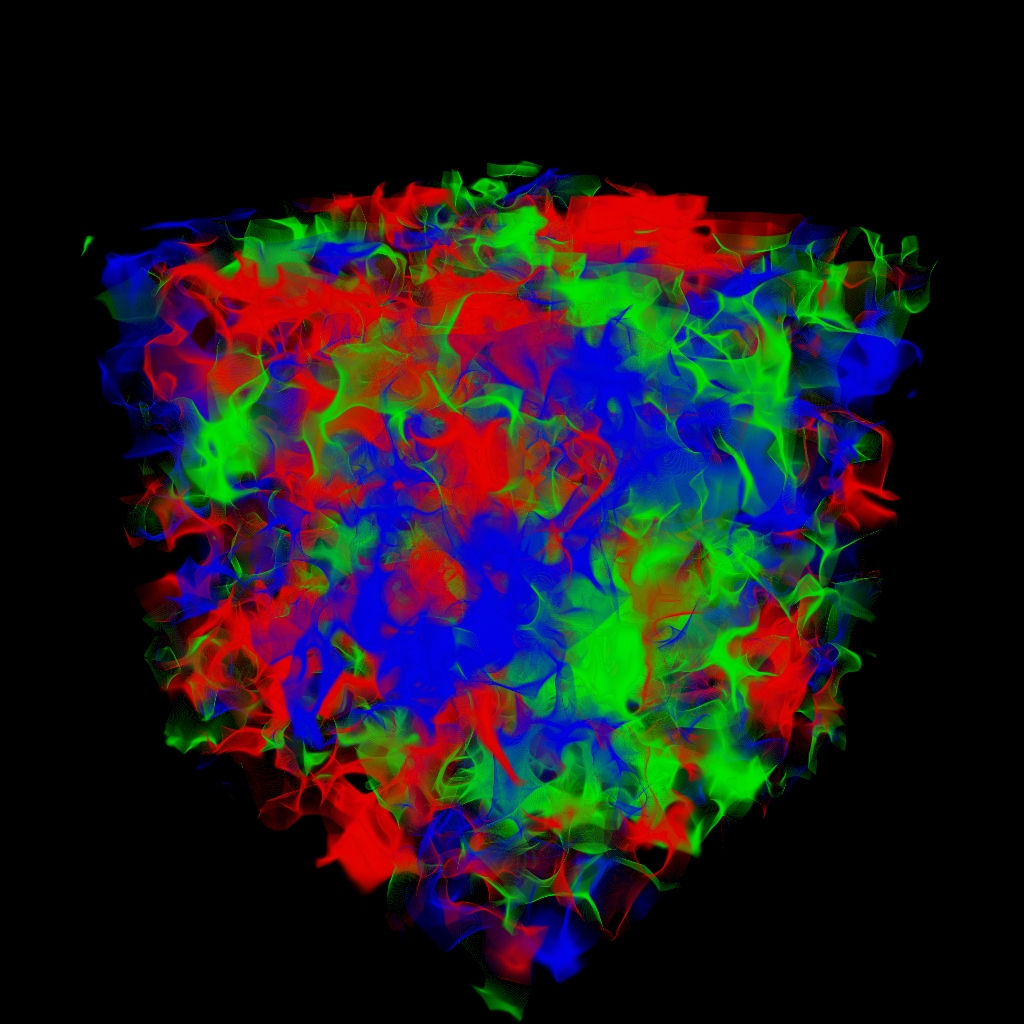}
    \label{fig:evolutionsmear1}
  }
  \qquad
  \subfloat[After one unit HMC trajectory]{
    \includegraphics[width=0.47\textwidth]{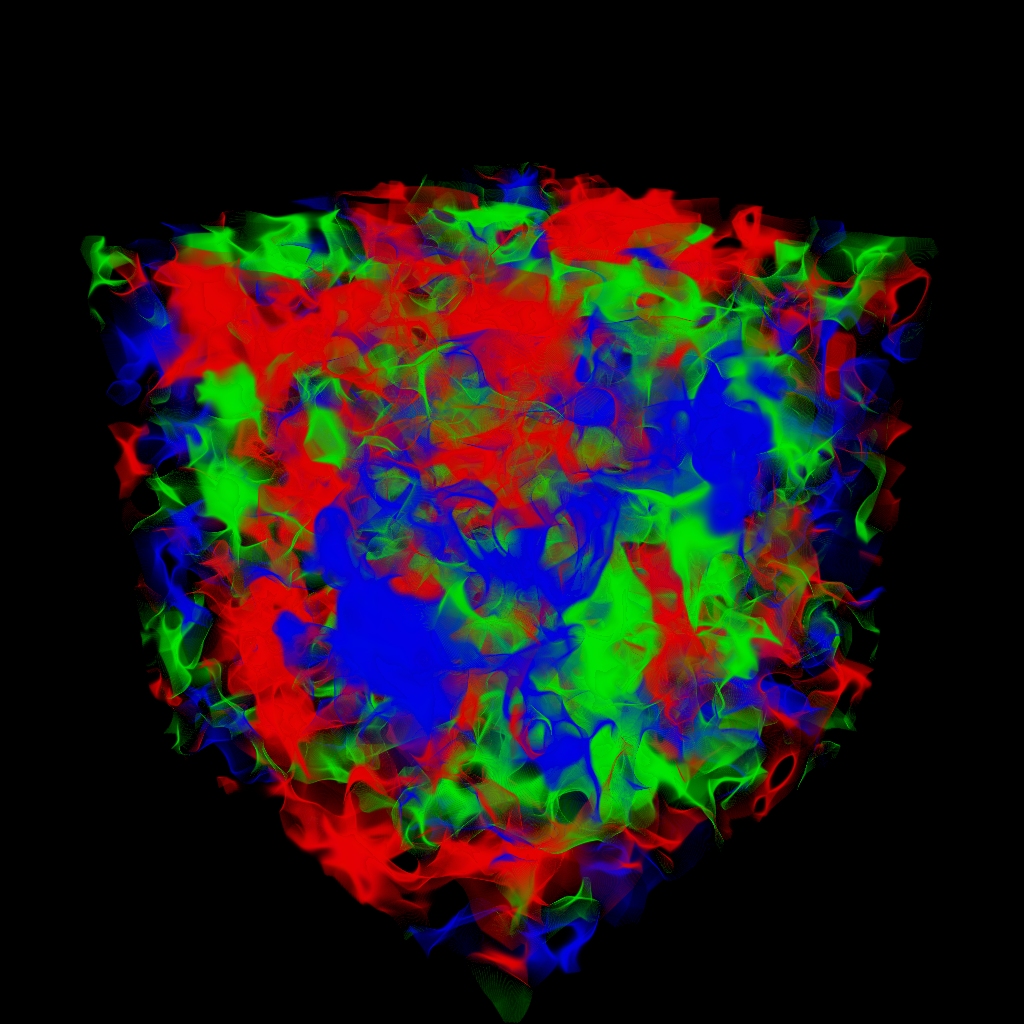}
    \label{fig:evolutionsmear2}
  }
  \caption{Evolution of center clusters with simulation time at
    \(T = 0.89(1) \, T_C\). Four sweeps of stout-link smearing are applied
    to the gauge links prior to calculating the Polyakov loops.}
  \label{fig:evolutionsmear}
\end{figure*}

\begin{figure*}[ptb]
  \includegraphics[width=\textwidth]{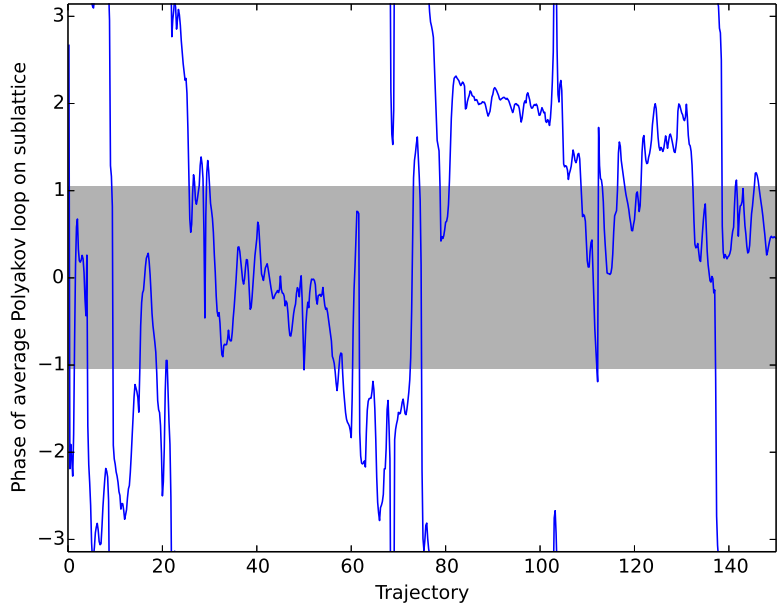}
  \caption{Evolution of phase of average Polyakov loop on \(6\times6\times6\)
    sublattice with simulation time at  \(T = 0.89(1) \, T_C\). Four sweeps of
    stout-link smearing are applied to the gauge links prior to calculating
    the Polyakov loops. Since the phase is \(2\pi\)-periodic, we see the graph
    wrapping around on the y-axis.}
  \label{fig:autocorrelation}
\end{figure*}

To investigate the larger scale behavior of the clusters,
we remove the small scale noise by
performing four sweeps of stout-link smearing \cite{Zhang:2009jf}
prior to calculating the Polyakov loops. The results are illustrated in
Fig.~\ref{fig:evolutionsmear}.
In these visualisations, we see that over the course of one trajectory, the
center clusters are slowly moving, with some change
around the boundaries of the clusters. Observing the evolution
of the center clusters in the corresponding animation,
we see correlations in the center clusters
persisting for approximately 5 seconds corresponding to 25 trajectories,
suggesting an approximate length for the autocorrelation time of the larger
scale structure of the clusters. After approximately 50
trajectories, the large scale structure of the loops has become completely
decorrelated. This is supported by
Fig.~\ref{fig:autocorrelation}, which shows the phase of the average Polyakov
loop on a small (\(6\times6\times6\)) subsection of the lattice. We can see
that the phase becomes completely decorrelated within 50 trajectories. This
supports our choice of separation for independent configurations.

\begin{figure*}[ptb]
  \subfloat[Thermalised configuration]{
    \includegraphics[width=0.47\textwidth]{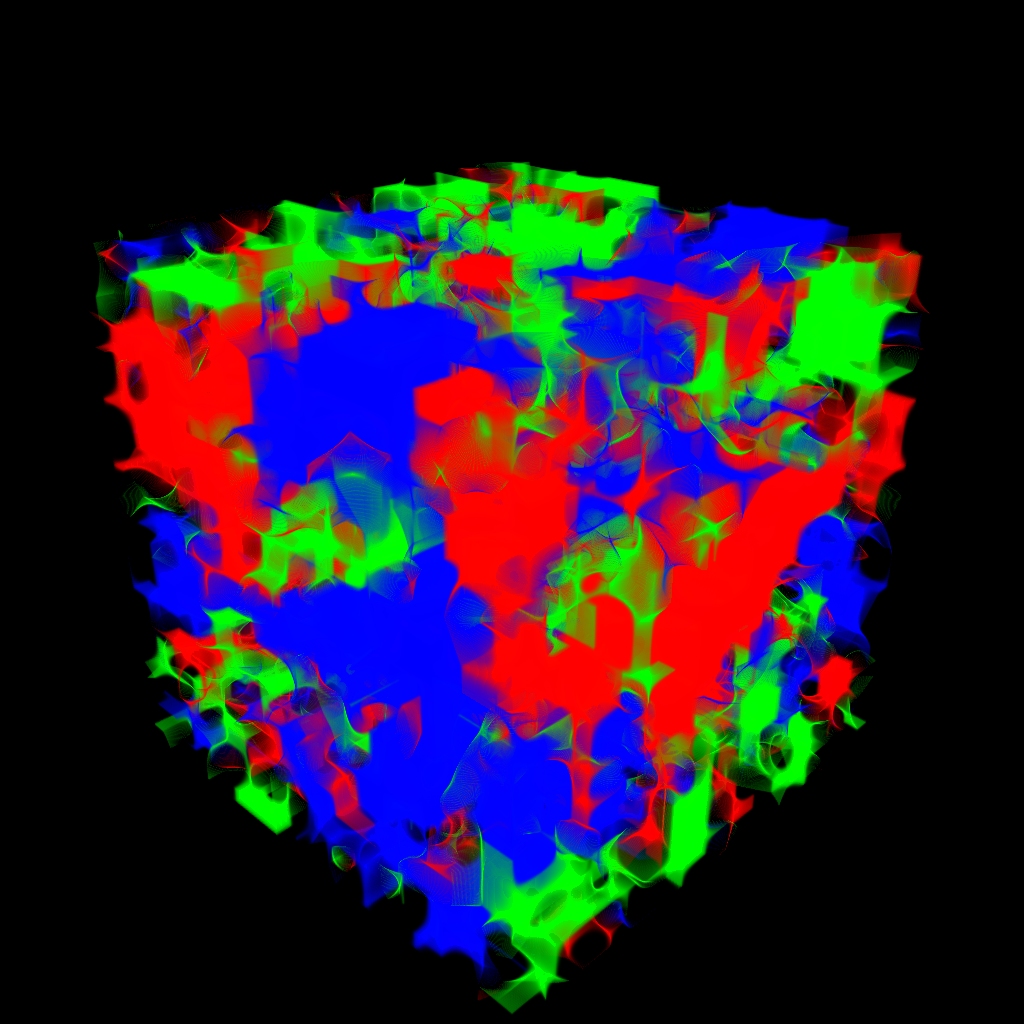}
    \label{fig:evolutionpotts1}
  }
  \qquad
  \subfloat[After 4096 single-site Metropolis updates]{
    \includegraphics[width=0.47\textwidth]{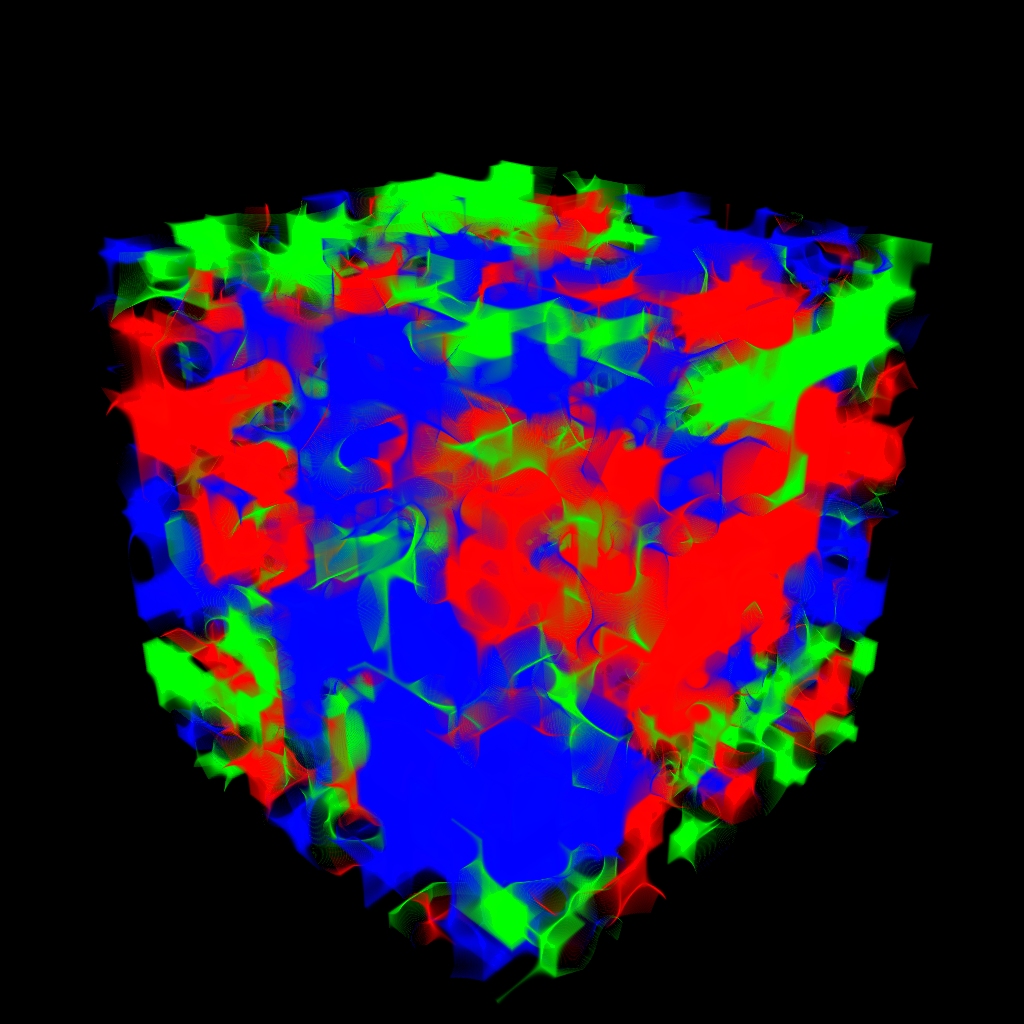}
    \label{fig:evolutionpotts2}
  }
  \caption{Evolution of spin-aligned domains in the three
    dimensional 3-state Potts model at \(\beta=0.55\) on a \(16^3\) lattice.}
  \label{fig:evolutionpotts}
\end{figure*}

The evolution observed in \(SU(3)\) gauge theory is consistent with the behavior
of spin-aligned domains in
the three dimensional 3-state Potts model \cite{Janke:1996qb} just below the
critical temperature under Metropolis-Hastings algorithm
simulations, as seen in Fig.~\ref{fig:evolutionpotts}.

We can also see that once a particular center phase
comes to dominate the space, that phase remains dominant for the rest of the
simulation. That is, in this first investigation of the behaviour of the Polyakov
loop under Monte-Carlo evolution, we find that the dominant phase is highly stable
under the process of HMC updates.

\subsection{Magnitude-Based Clusters\label{sec:results:mag}}
\begin{figure*}[ptb]
  \subfloat[Opacity based on \(\phi(\vec{x})\)]{
    \includegraphics[width=0.47\textwidth]{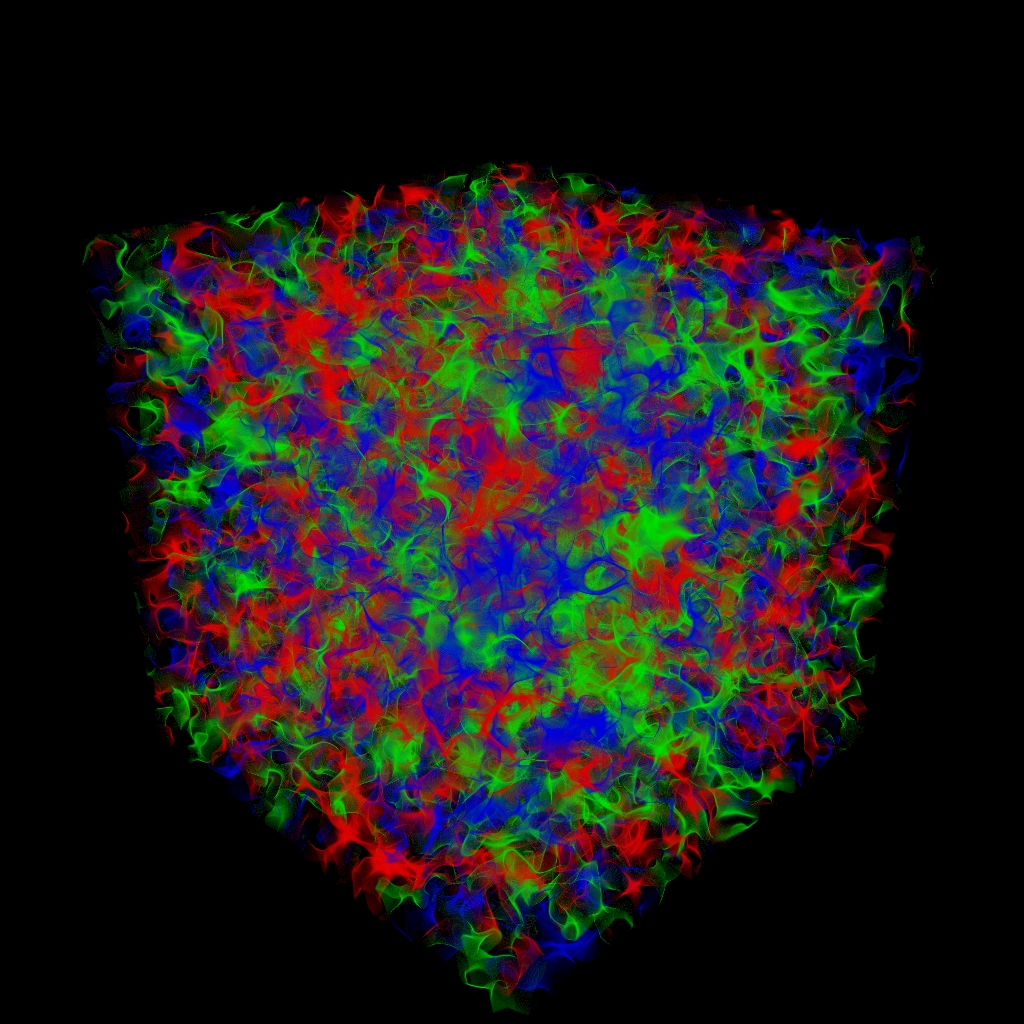}
    \label{fig:nonmod1.0}
  }
  \qquad
  \subfloat[Opacity based on \(\rho(\vec{x})\)]{
    \includegraphics[width=0.47\textwidth]{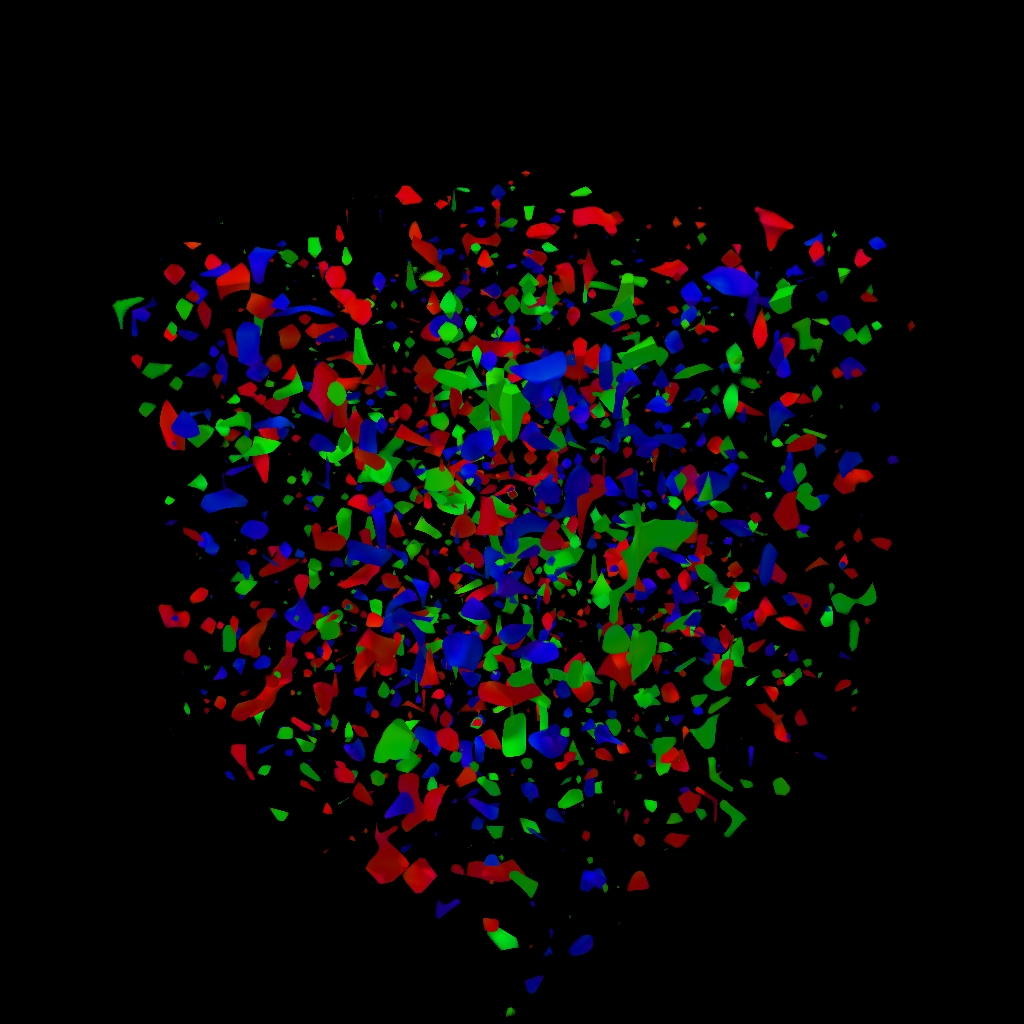}
    \label{fig:mod1.0}
  }
  \caption{Comparison of complex phase \(\phi(\vec{x})\) and
    magnitude \(\rho(\vec{x})\) clusters at \(T = 0.89(1) \, T_C\).}
  \label{fig:argvsmod1.0}
\end{figure*}

If we use the alternate rendering style based on the magnitude \(\rho(x)\), we
can see that peaks in the magnitude
lie approximately within the center clusters, as shown in
Fig.~\ref{fig:argvsmod1.0}.
The magnitude peaks are each colored corresponding to a single center phase
and they appear to line up
with peaks in the corresponding phase based visualisation. However it is not
necessarily clear that each
center cluster corresponds to a peak in the modulus, \(\rho(\vec{x})\).

\begin{figure*}[ptb]
  \subfloat[Clusters based on \(\phi(\vec{x})\)]{
    \includegraphics[width=0.47\textwidth]{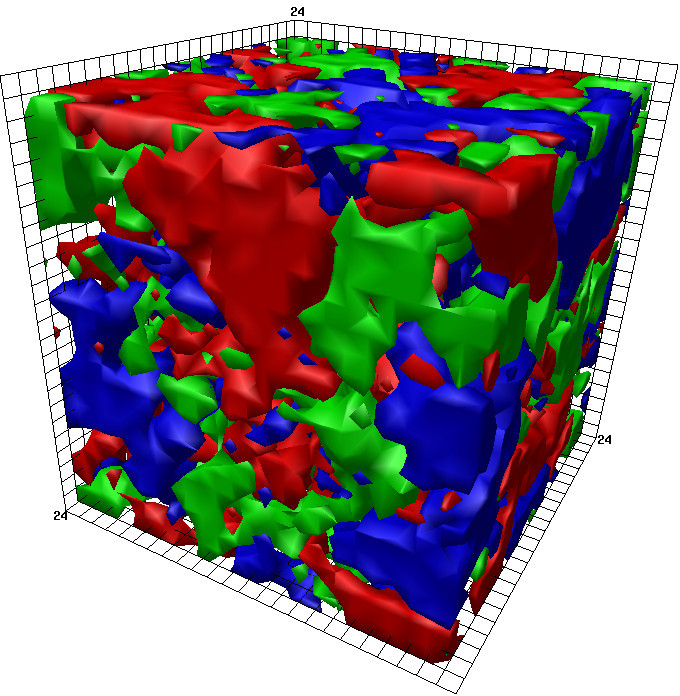}
    \label{fig:nonmodsmear0.2}
  }
  \qquad
  \subfloat[Clusters based on \(\rho(\vec{x})\)]{
    \includegraphics[width=0.47\textwidth]{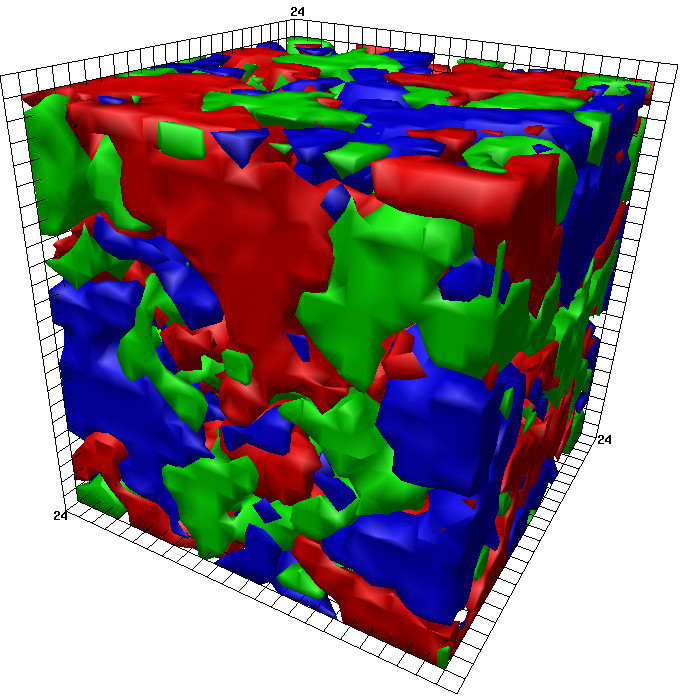}
    \label{fig:modsmear0.2}
  }
  \caption{Comparison of phase \(\phi(\vec{x})\) and magnitude \(\rho(\vec{x})\)
   clusters at \(T = 0.89(1) \, T_C\) after four sweeps of stout-link smearing using the isovolume
   renderer. The rendering thresholds for \(\phi(\vec{x})\) and \(\rho(\vec{x})\) are
   0.5 and 0.2 respectively.}
  \label{fig:argvsmodsmear0.2}
\end{figure*}

\begin{figure*}[ptb]
  \subfloat[Clusters based on \(\phi(\vec{x})\)]{
    \includegraphics[width=0.47\textwidth]{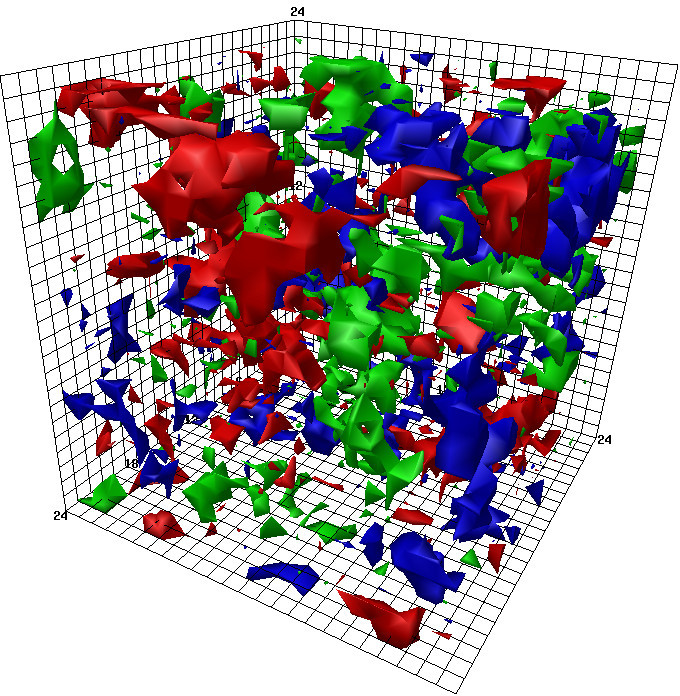}
    \label{fig:nonmodsmear0.5}
  }
  \qquad
  \subfloat[Clusters based on \(\rho(\vec{x})\)]{
    \includegraphics[width=0.47\textwidth]{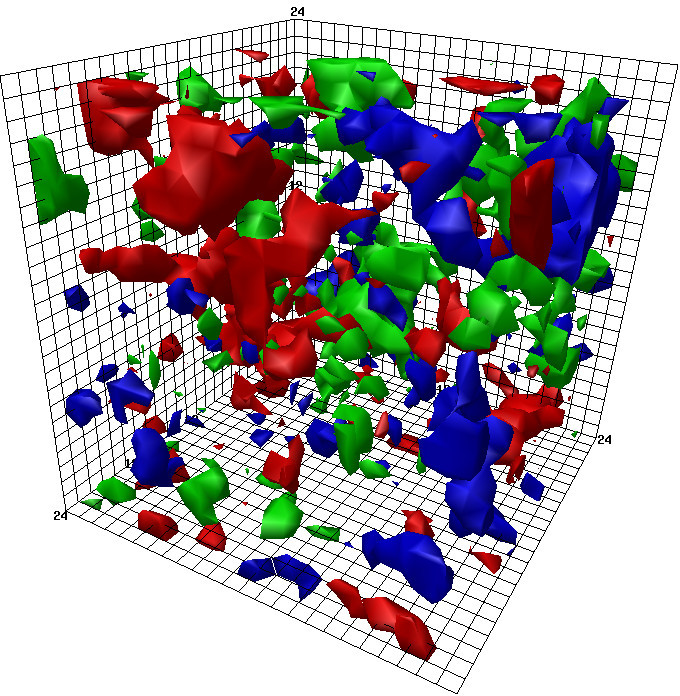}
    \label{fig:modsmear0.5}
  }
  \caption{Comparison of phase \(\phi(\vec{x})\) and magnitude \(\rho(\vec{x})\)
   clusters at \(T = 0.89(1) \, T_C\) after four sweeps of stout-link smearing, using the isovolume
   renderer, cut much closer into the peaks than Fig.~\ref{fig:argvsmodsmear0.2}. Here, the rendering
   thresholds for \(\phi(\vec{x})\) and \(\rho(\vec{x})\) are 0.9 and 0.5 respectively.}
  \label{fig:argvsmodsmear0.5}
\end{figure*}

\begin{figure*}[ptb]
  \includegraphics[width=\textwidth]{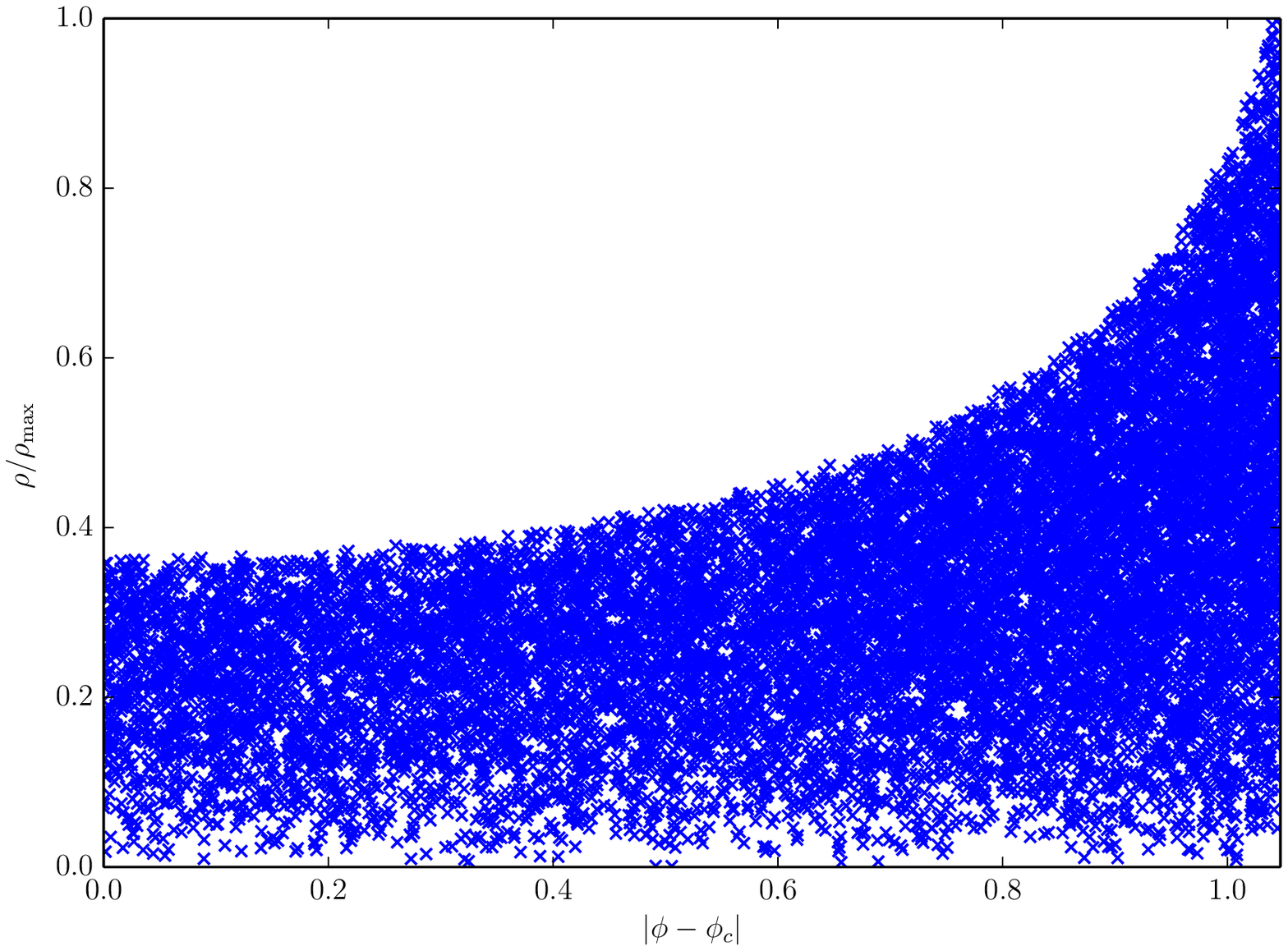}
  \caption{Scatter plot of center phase proximity and modulus, showing a
      clear correlation between proximity to a center phase and maxima of the modulus.}
  \label{fig:phasemodscatter}
\end{figure*}

In order to look more closely at the correlation between these clusters, we
again perform four sweeps of stout-link smearing,
making it easier to observe larger scale structures, and render
both the phase and magnitude based clusters using an isovolume rendering
developed in AVS/Express~\cite{AVSExpress} that makes the boundaries of the
clusters more clear. When we do this, it
becomes clear that there is an approximate one-to-one
relationship between the peaks in the magnitude \(\rho(\vec{x})\) and the
proximity of \(\phi(\vec{x})\) to a center phase. The measures
\(|\phi(\vec{x})-\phi_c(\vec{x})|/(\pi/3)\) and \(\rho(x) / \rho_{\mathrm{max}}\) are
illustrated in Figs.~\ref{fig:argvsmodsmear0.2} and \ref{fig:argvsmodsmear0.5}.
We can also observe this correlation by looking at a scatter plot of
\(\rho(x) / \rho_{\mathrm{max}}\) vs \(|\phi(\vec{x})-\phi_c(\vec{x})|\)
as in Fig.~\ref{fig:phasemodscatter}. This observed correlation of
\(\rho(\vec{x})\) becoming small as \(\phi(\vec{x})\) moves away from a center
phase is the first direct confirmation of the underlying assumption of
Ref.~\cite{Asakawa:2012yv}, which links the center domain walls to unanticipated
phenomena observed at RHIC \cite{Muller:2006ee} and the LHC 
\cite{Chatrchyan:2011sx,Aad:2010bu,Tonjes:2011zz}.

These peaks in the magnitude of the local Polyakov loops mean that the free
energy of a quark-antiquark pair
is minimised in the core of a center cluster. Thus we have a confining potential
with local minima at the cores of
center clusters. Below the critical temperature the peaks are sharp so the
gradient of the potential is
steep, resulting in a strong restoring force confining the quarks to the core of
the cluster. Above the critical
temperature, the peaks become much
broader so the gradient is significantly smaller
and the restoring force is greatly reduced.

\subsection{Center Clusters\label{sec:results:clusters}}
The local Polyakov loop \(L(\vec{x})\) is a complex number at each spatial
lattice site \(\vec{x}\). We can express it in complex polar form,
\[L(\vec{x}) = \rho(\vec{x}) \, e^{i\phi(\vec{x})}\, ,\]
where \(\rho(\vec{x}), \phi(\vec{x}) \in \mathbb{R}\), \(\rho(\vec{x}) \geq 0\),
and \(-\pi < \phi(\vec{x}) \leq \pi\).

In order to observe the structure of individual clusters, we
use the definition by Gattringer and Schmidt \cite{Gattringer:2010ug}. Two
neighboring points \(\vec{x}\) and \(\vec{y}\) belong to the same cluster iff
\(n(\vec{x}) = n(\vec{y})\), where the sector number \(n(\vec{x})\) is defined to be
\[n(\vec{x}) := \left\{
\begin{array}{lr}
 -1 & \mathrm{for} \, \phi(\vec{x}) \in [-\pi + \delta, -\frac{\pi}{3} - \delta],\\
  0 & \mathrm{for} \, \phi(\vec{x}) \in [-\frac{\pi}{3} + \delta, \frac{\pi}{3} - \delta],\\
  1 & \mathrm{for} \, \phi(\vec{x}) \in [\frac{\pi}{3} + \delta, \pi - \delta].
\end{array}
\right.\]

\begin{figure*}[ptb]
  \subfloat{
    \includegraphics[width=0.47\textwidth]{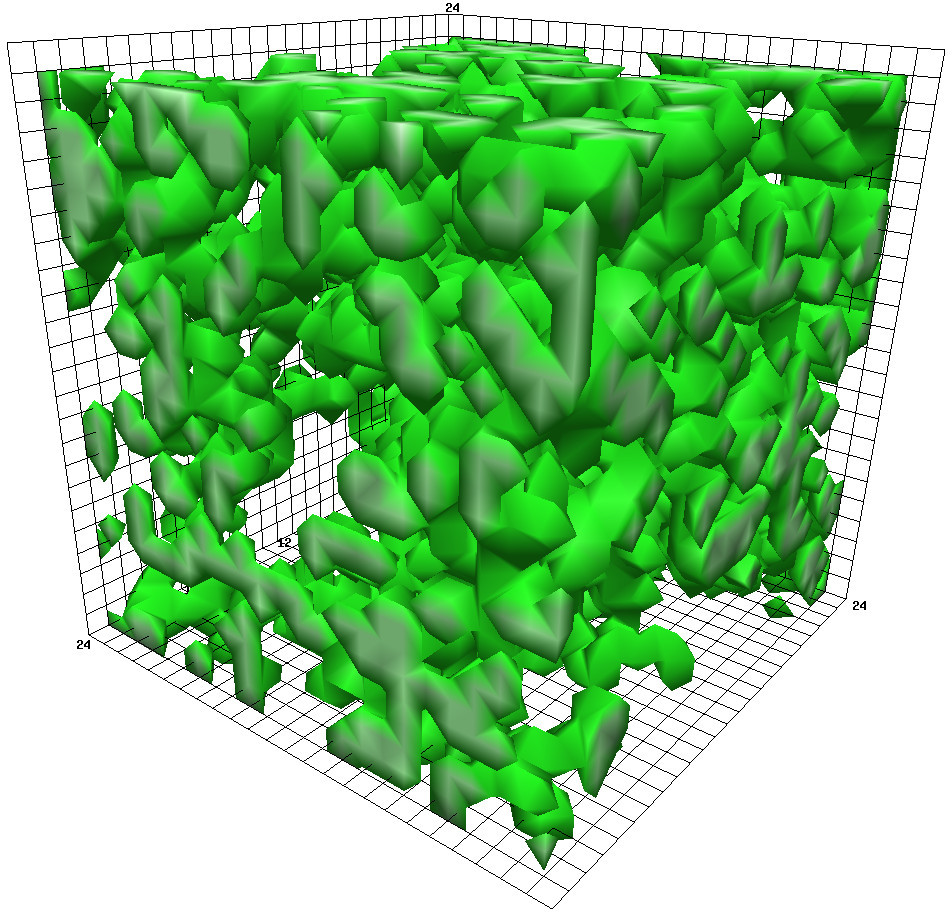}
  }
  \qquad
  \subfloat{
    \includegraphics[width=0.47\textwidth]{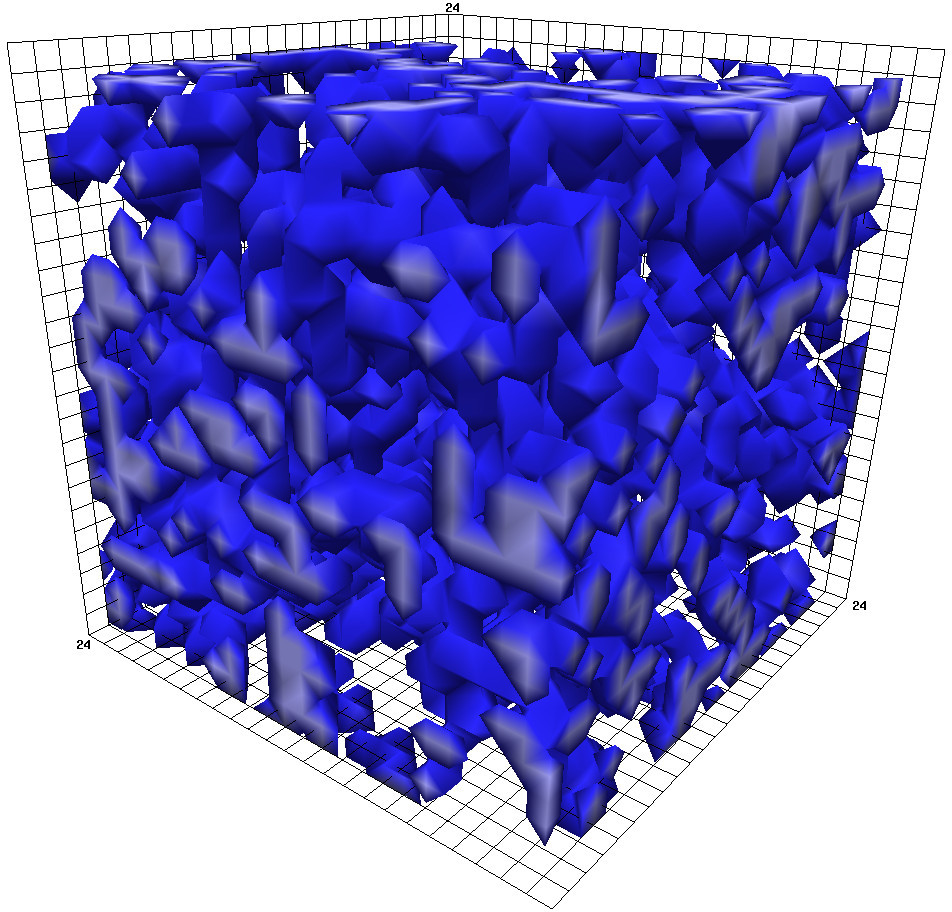}
  }
  \caption{Two of the three percolating clusters in a representative
    \(24^3 \times 8\) gauge field configuration at \(T = 0.89(1) \, T_C\)
    when using a cut parameter of zero.}
  \label{fig:percolation}
\end{figure*}

If we include every site in a cluster (i.e. we set the cut
parameter \(\delta\) to zero), then below the critical temperature each sector is
equally occupied so our clustering is comparable to random site percolation
theory with an occupation probability of \(p \approx 0.3333\). This is above the
critical percolation probability of \(p_c = 0.3116\)
\cite{doi:10.1142/S0129183198000261}. Thus, with a cut
parameter of zero, we expect to see at least one percolating cluster (i.e. a
cluster that has at least one site in each of the \(3N_s\) spatial planes) for
each phase. Thus, if we define the clusters in this way, due to the nature of
three dimensional space they are not localized, and instead extend over the
entire lattice, as seen in Fig.~\ref{fig:percolation}.

\begin{figure*}[ptb]
  \subfloat{
    \includegraphics[width=0.47\textwidth]{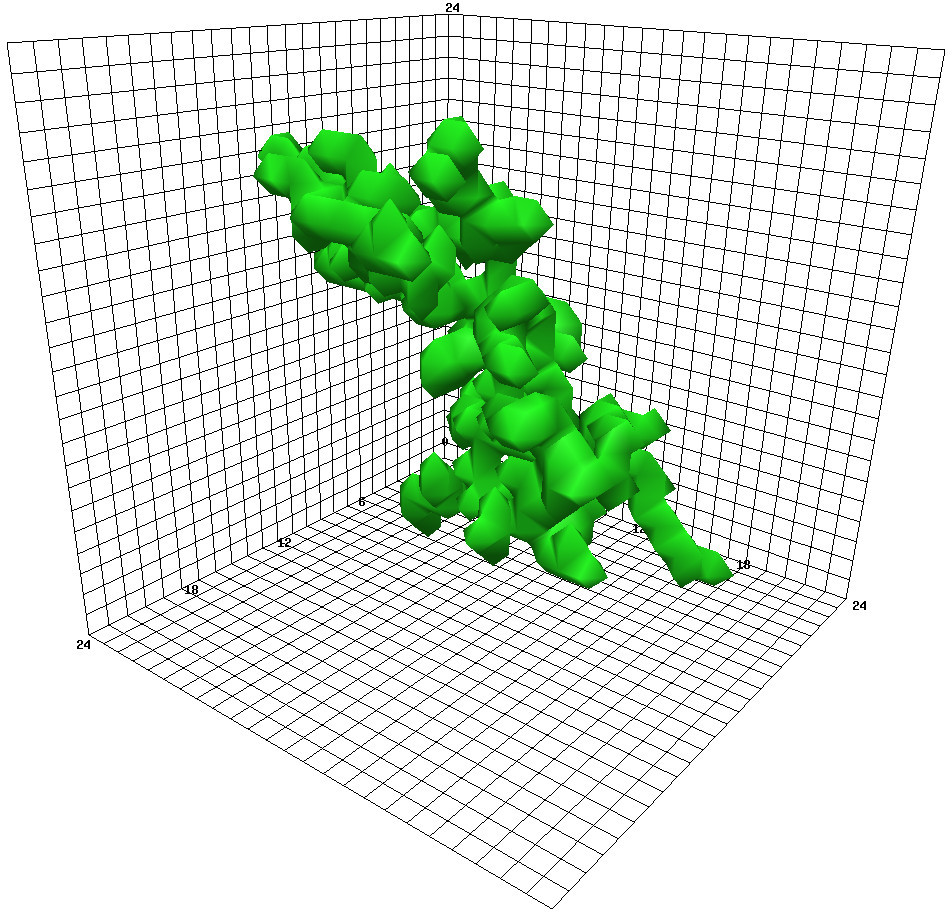}
  }
  \qquad
  \subfloat{
    \includegraphics[width=0.47\textwidth]{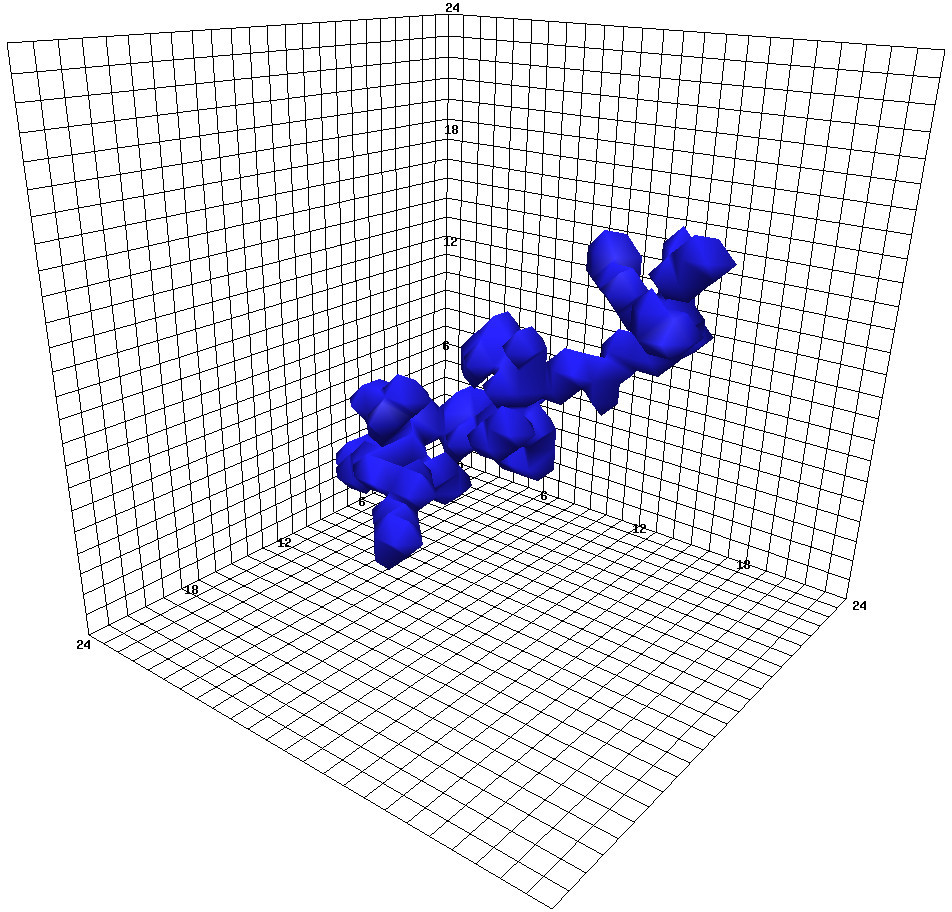}
  }
  \caption{One of the non-percolating clusters broken off from each of the clusters
    in Fig.~\ref{fig:percolation} when the cut parameter is increased to 0.3.}
  \label{fig:cut}
\end{figure*}

In order to study localized clusters below the critical temperature, we instead
introduce a cut parameter as in Ref. \cite{Gattringer:2010ug}. This results in
smaller, localized clusters scattered across the lattice. Fig.~\ref{fig:cut}
illustrates results for \(\delta = 0.3\). We can see that these clusters
have a one dimensional, finger-like quality to their structure and this supports
the determination by Endrodi, Gattringer and Schadler \cite{Endrodi:2014yaa}
that the center clusters have a fractal dimension of \(D \approx 1.4 - 1.7\).

As illustrated in Fig.\ref{fig:phasemodscatter}, the selection of a cut in
\(\phi(\vec{x})\) identifies a subset of points where \(\rho(\vec{x})\) is small
and attributes the points to a domain wall of finite thickness.

Just as the expectation value of \(L\) is related to the free energy of a static
quark relative to the vacuum by
\[\left<L(\vec{x})\right> \propto \exp(-F_q/T)\, ,\]
the correlation function \(\left<L(\vec{x})L(\vec{y})^\dagger\right>\) is related
to the free energy of a static quark-antiquark
pair (a meson) separated by \(\vec{x}-\vec{y}\) \cite{Svetitsky:1982gs}:
\[\left<L(\vec{x})L(\vec{y})^\dagger\right> \propto \exp(-F_{q\bar{q}}(\vec{x}
    -\vec{y})/T)\,.\]
Thus:
\begin{enumerate}
  \item If \(\left<L(\vec{x})L(\vec{y})^\dagger\right> \ne 0\), then
    \(F_{q\bar{q}}(\vec{x}-\vec{y})\) is finite.
  \item If \(\left<L(\vec{x})L(\vec{y})^\dagger\right> = 0\), then
    \(F_{q\bar{q}}(\vec{x}-\vec{y})\) is infinite.
\end{enumerate}
If we consider these smaller, localized center clusters, we can then conclude
the following:
\begin{enumerate}
  \item If \(\vec{x}\) and \(\vec{y}\) lie within a single cluster,
    \(\phi(\vec{x}) \approx \phi(\vec{y})\) and
    \begin{align*}
      \left<L(\vec{x})L(\vec{y})^\dagger\right> &= \left<\rho(\vec{x})
        \rho(\vec{y})e^{i(\phi\left(\vec{x})-\phi(\vec{y})\right)}\right> \\
        &\approx \left<\rho(\vec{x})\rho(\vec{y})\right> \\
        &\ne 0\, .
    \end{align*}
    That is, the phases at \(\vec{x}\) and \(\vec{y}\) cancel out, and the
    correlation function is non-zero. Thus,
    \(F_{q\bar{q}}(\vec{x}-\vec{y})\) is finite and quark-antiquark pairs can be
    separated within a cluster.
  \item If \(\vec{x}\) and \(\vec{y}\) lie within different clusters,
    \(\phi(\vec{x}) \ne \phi(\vec{y})\) in
    general and thus there is no correlation between the phases.
    Due to the symmetry of the distribution of the phases between
    clusters, the three center phases are equally present. Across an ensemble
    they average to zero, giving
    \begin{align*}
      \left<L(\vec{x})L(\vec{y})^\dagger\right> &= 0
    \end{align*}
    Thus, \(F_{q\bar{q}}(\vec{x}-\vec{y})\) is infinite and
    quark-antiquark pairs cannot be separated
    across cluster boundaries.
\end{enumerate}

\begin{figure*}[ptb]
  \includegraphics[width=0.9\textwidth]{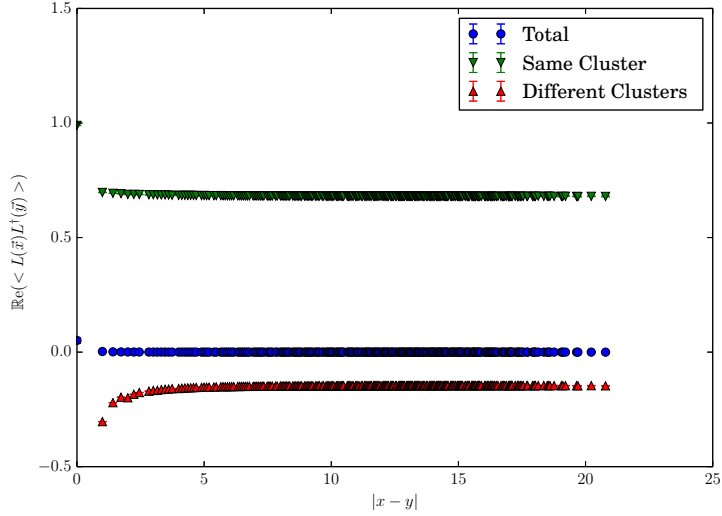}
  \caption{Comparison of real component of correlation function for points in the same
    and in different clusters and averaged across all points. Generated from \(24^3 \times 8\) lattice
    at \(T = 0.89(1) \, T_C\) with a cut parameter of \(0.0\).}
  \label{fig:recf0.0}
\end{figure*}

\begin{figure*}[ptb]
  \includegraphics[width=0.9\textwidth]{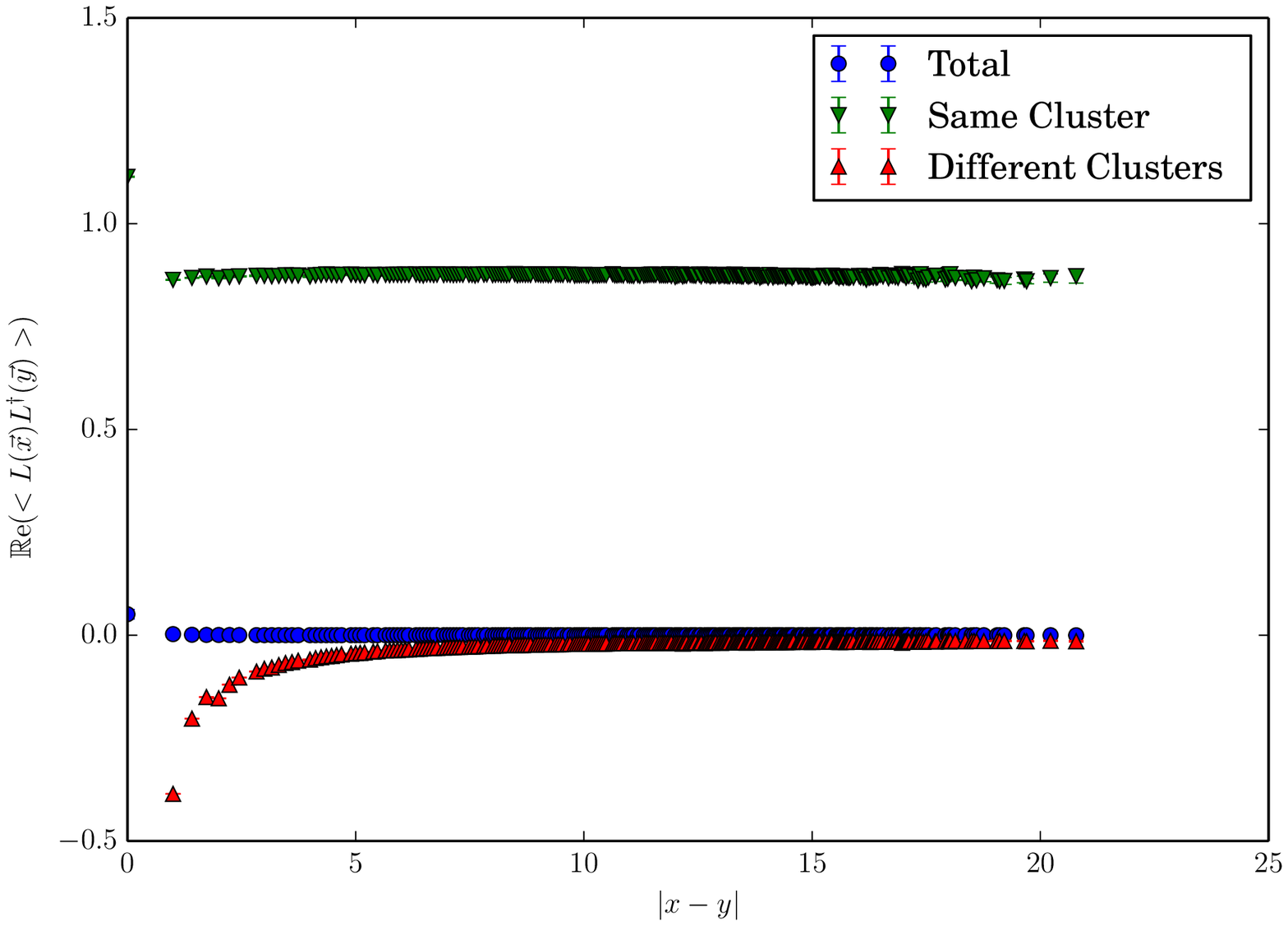}
  \caption{Comparison of real component of correlation function for points in the same
    and in different clusters and averaged across all points. Generated from \(24^3 \times 8\) lattice
    at \(T = 0.89(1) \, T_C\) with a cut parameter of \(0.3\).}
  \label{fig:recf0.3}
\end{figure*}

\begin{figure*}[ptb]
  \includegraphics[width=0.9\textwidth]{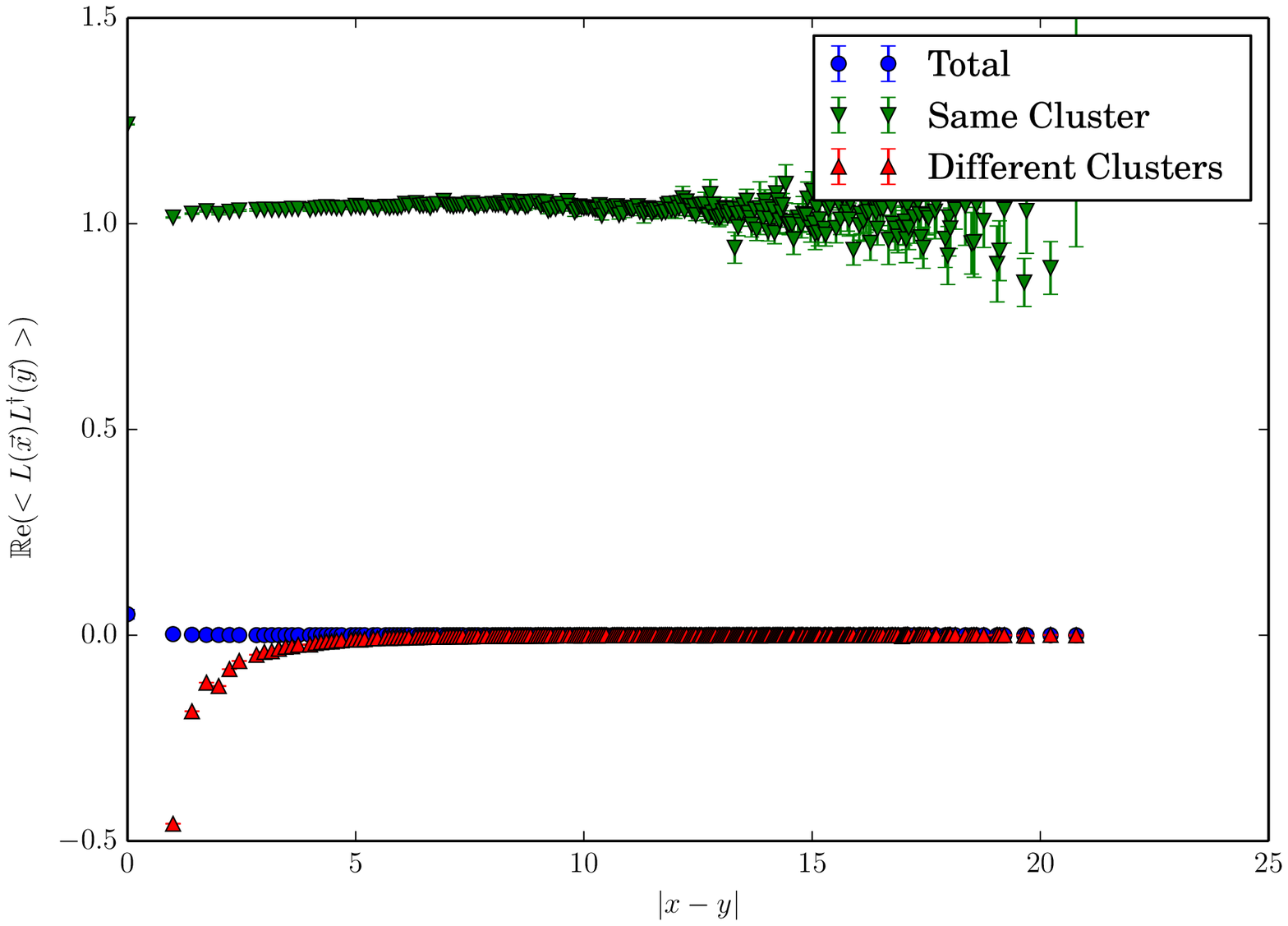}
  \caption{Comparison of real component of correlation function for points in the same
    and in different clusters and averaged across all points. Generated from \(24^3 \times 8\) lattice
    at \(T = 0.89(1) \, T_C\) with a cut parameter of \(0.5\).}
  \label{fig:recf0.5}
\end{figure*}

\begin{figure*}[ptb]
  \includegraphics[width=\textwidth]{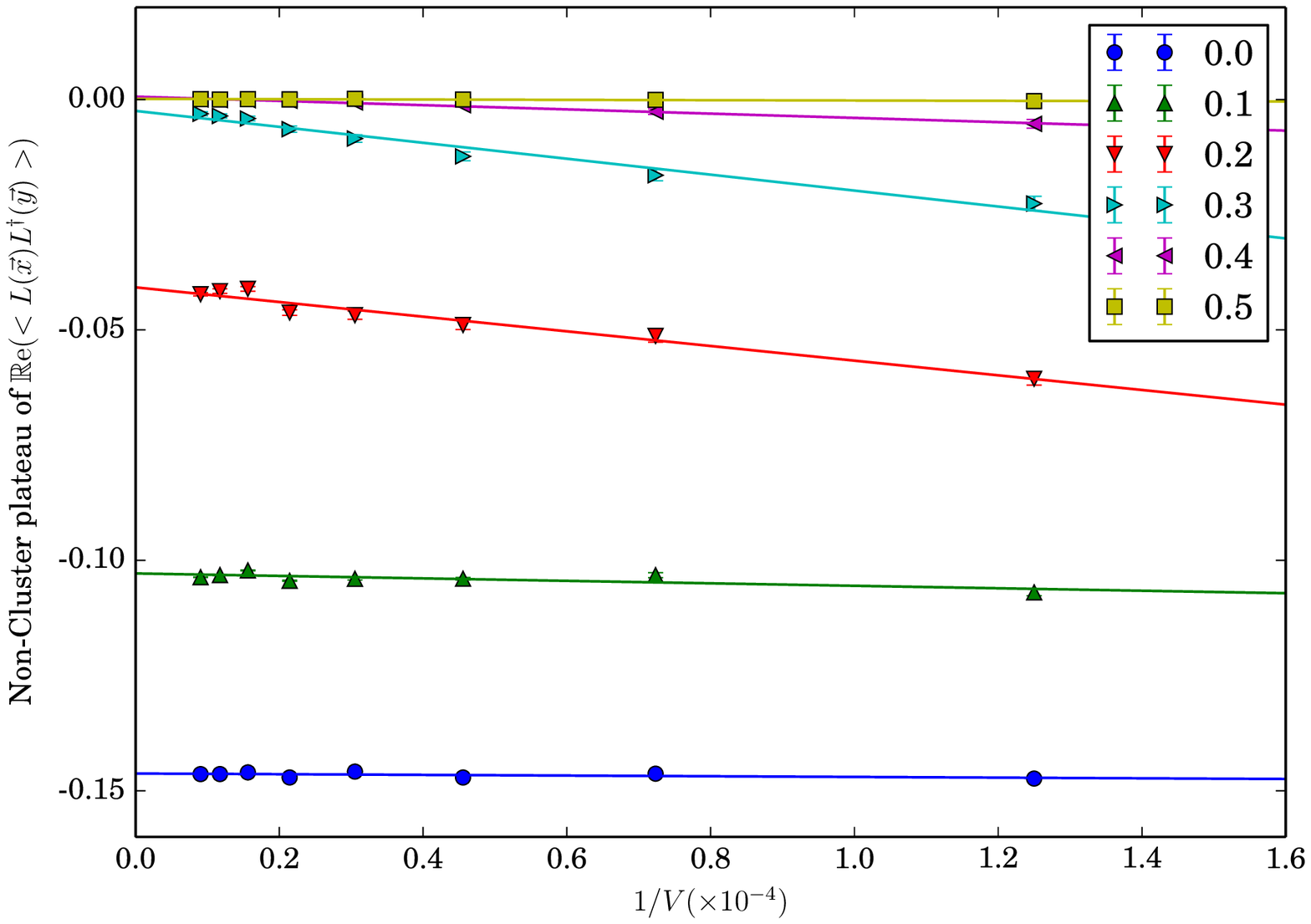}
  \caption{Volume and cut parameter dependence of the plateau in the correlation function
    for different clusters at \(T = 0.89(1) \, T_C\). Cut parameters are indicated in the
    legend.}
  \label{fig:voldep}
\end{figure*}

We can observe that this is approximately the case for sufficiently large
\(\left|\vec{x} - \vec{y}\right|\) in Figs.~\ref{fig:recf0.0}-\ref{fig:recf0.5}, produced from an
ensemble of 100 independent gauge field configurations, which show that the real component
of the correlator is nonzero and positive for points in the same cluster and plateaus to a small
negative value for points in different clusters. The imaginary component is negligible. As
can be seen from Fig.~\ref{fig:voldep}, the nonzero negative value of the plateau is a finite volume
effect that goes to zero in the infinite volume limit for sufficiently large values of the
cut parameter. This effect
results from the cluster at one point (say \(\vec{x}\)) extending to the neighborhood of
\(\vec{y}\) and disturbing the even distribution of clusters between the three phases. In order
for the correlation function to vanish, we need a cut parameter that is sufficiently large that
the clusters have a finite size and a volume significantly larger than that size. This allows us to
access an even distribution between the three phases so that it can average to zero.

Thus, the cluster size has a physical significance, governing the confining
scale of the theory and the size of mesons. Below the critical temperature,
when (given a sufficiently large cut parameter),
the clusters have a limited finite size, this results in confinement. From a model
perspective, this scale governs the size of the quark core
of hadrons which is dressed by the mesonic cloud. Above the critical
temperature, as a single cluster grows to encompass the
entire space (regardless of the cut used), the quarks become deconfined
\cite{Gattringer:2010ug}.

\section{Conclusion\label{sec:conclusion}}
We simulated SU(3) Yang-Mills theory on anisotropic \(24^3\times8\) lattices
with the Iwasaki gauge action, considering renormalised anisotropy ranging
from \(\xi \approx 0.99\) to \(\xi \approx 1.80\) with a spatial lattice spacing
of \(a_s \approx 0.1\). We explored temperatures ranging from \(0.89(1)\,T_C\)
to \(1.61(3)\,T_C\). Our focus is on the structure and evolution of center
clusters associated with Polyakov loops.

In doing so, we developed a volume rendering program that correctly deals with
the interpolation of three dimensional complex fields such as the
local Polyakov loop, and clearly displays their phase (and/or absolute value).
This allows us to observe the evolution of center clusters with HMC simulation
time.

For the first time, we were able to reveal the evolution of
center clusters as they transitioned from the confined to the deconfined phase.
The cluster behavior is consistent with
that of spin aligned domains in the three dimensional 3-state Potts model. This
supports the idea that the phase transition in QCD is comparable
to that of a three state spin system.

We also observe an approximate one to one correspondence between peaks in the
magnitude of local Polyakov loops and the locations of center clusters defined
by the proximity of the phase of the Polyakov loops to a center phase. Gaps
between the clusters illustrate the magnitude of the Polyakov loop is
suppressed within the domain walls. This observation supports the underlying
assumption of Ref.~\cite{Asakawa:2012yv}, linking
the center domain walls to phenomena observed at RHIC
\cite{Muller:2006ee} and the LHC \cite{Chatrchyan:2011sx,Aad:2010bu,
Tonjes:2011zz}.

The creation of a domain wall of finite thickness between clusters through the
cut parameter \(\delta\) \cite{Gattringer:2010ug} produces clusters in which the
quark-antiquark correlation function \(\left<L(\vec{x})L^\dagger(\vec{y})\right>\)
vanishes when both quarks do not reside in the same cluster, setting a scale
for confinement.

These peaks in the magnitude are sharp
below the critical temperature. Through considerations of the free energy of
multi-quark systems, this results in a minimum of the quark-antiquark potential
in the core of each center cluster with a steep
gradient resulting in a strong restoring force confining the quarks within the
cluster. Above the critical temperature the peak structure becomes smooth
and the region covered by the domain becomes large. When averaged over an
ensemble, any remaining fluctuations in the Polyakov loops are smoothed out and
the restoring force becomes negligible. In this way, quarks become deconfined.

\appendix

\section{Visualisation\label{sec:visualisation}}
\subsection{Algorithm\label{sec:visualisation:alg}}
In order to perform the visualisation, we use a ray-traced volume renderer.
For each pixel in the final image, a single ray is traced out directly away
from the viewer through the volume to be
rendered, accruing color and opacity based on the volumetric data. 

Given a RGB (red, green, blue) color vector \(C^{vol}(\vec{x})\) and an opacity
\(\alpha^{vol}(\vec{x})\) at every point \(\vec{x}\) in the volume,
we accrue color and opacity along a ray \(\vec{x}(z)\) (\(0 \leq z \leq 1\),
where \(\vec{x}(0)\) is the point where the ray
enters the volume and \(\vec{x}(1)\) the point where it exits) by the
differential equations,
\begin{align*}
  \frac{d\alpha^{ray}(z)}{dz} &= \left(1-\alpha^{ray}(z)\right)\alpha^{vol}
    \left(\vec{x}(z)\right)\, , \\
  \frac{dC^{ray}(z)}{dz} &= \left(1-\alpha^{ray}(z)\right)C^{vol}\left(\vec{x}(z)
    \right)\, .
\end{align*}

To solve these differential equations, we use Euler's method, with a finite step
size \(\Delta z\)
\cite{Levoy1990245}:
\begin{align*}
  \alpha^{ray}_{n+1} &= \alpha^{ray}_n + \Delta z \left(1-\alpha^{ray}_n\right)
  \alpha^{vol}\left(\vec{x}(z_n)\right)\, , \\
  C^{ray}_{n+1} &= C^{ray}_n + \Delta z \left(1-\alpha^{ray}_n\right)C^{vol}\left(\vec{x}(z_n)\right)\, .
\end{align*}

In order to perform these calculations, we use OpenGL \cite{OpenGL}, a 2D and 3D
graphics API that allows us to leverage
the powerful hardware available in modern GPUs which is designed specifically
for rendering graphics. OpenGL
provides a flexible graphics processing pipeline which for our purposes consists
of a vertex shader followed by a fragment
shader. The vertex shader takes in information about the position and shape of
three dimensional objects to be rendered
and transforms them into the two dimensional space of the screen. The fragment
shader runs once for each pixel on the screen,
taking information about the polygon visible at that point from the vertex
shader and determining the color the pixel should be.

In our particular case, we adapt a technique by Kruger and Westermann
\cite{Kruger:2003} which involves repeatedly rendering a
single cube with a sequence of different shader pairs. The vertex shader is the
same every time and performs a simple transformation on the
cube and calculates the mapping between points on the surface of the cube and
points in the volume data that is being rendered.
We then have three different fragment shaders that are run in sequence to
produce the desired output.  A feature that we make
extensive use of in order to store interim data is rendering to a framebuffer,
an image in memory that serves as a
virtual screen, allowing us to store the result of one fragment shader and then
use it in a later shading run.

The first fragment shader is run with only the outside faces of the cube
visible, so the vertex shader gives the
coordinates of the point a ray cast through the current pixel would enter the
volume. We
store these directly in the red, green, and blue channels of a framebuffer. We
then run the second shader with
only the inside faces of the cube visible, so the vertex shader gives the
coordinates of the point the ray
would leave the volume. We then access the entry coordinates from the
framebuffer and calculate the direction and length
of the ray inside the volume and store them in the red, green, blue, and alpha
channels of a new framebuffer.

We can then use a more complicated fragment shader to preform the integration,
indexing into a 3D texture containing the volume
data and calculating \(C^{vol}(\vec{x})\) and \(\alpha^{vol}(\vec{x})\) at each
step. By setting the interpolation mode on the texture, we
can tell OpenGL to automatically and efficiently perform trilinear interpolation
on the data.

By transforming the cube, we can transform the volume being rendered. Thus we
perform standard OpenGL
model/view and projection transformations on the cube to place the volume in the
center of the screen with 
perspective and continuously rotate it so that it is possible to see all sides
of the volume and observe
the 3D structure it contains.

We can then load the pixel data produced by the GPU back into main memory and
convert it into a range
of formats for later use. In particular, we use DevIL\cite{DevIL} to convert a
single frame into a static
image, or we use FFmpeg\cite{FFmpeg} to combine a sequence of frames
into a video.

\subsection{Optimization\label{sec:visualisation:opt}}
Implementing this algorithm naively is rather inefficient for all but the most
sparse data sets, as much of the volume is obscured
by opaque or nearly opaque regions and has little to no effect on the final
image. In order to eliminate this
inefficiency, we introduce early ray termination, that is we stop integrating
rays once they reach a
certain opacity threshold.

We do this by introducing another fragment shader that writes to the depth
buffer without changing the color
or opacity. The depth buffer is a special texture used by OpenGL to determine
what geometry should be obscured
by other geometry. If the opacity has reached a certain threshold, our shader
writes the minimum
possible value to the depth buffer, effectively terminating the ray.

This shader is then interleaved with the integrating shader, which writes to a
framebuffer to store its interim
result. This method is also used to stop integrating rays that have left the
volume, simply by setting their opacity to \(1\) if
they are outside the volume (equivalent to hitting a solid black backdrop).

In order to maximize efficiency, the integrating shader performs batches of
several steps at a time, starting from zero
opacity and color. The result is then appended to the previously calculated
integration by using OpenGL blending with
the blending mode set to
\[\alpha \to \alpha + (1-\alpha) \alpha_{new}\, ,\]
\[C \to C + (1-\alpha) C_{new}\, .\]

This tells OpenGL how to mix the new color produced by the fragment shader
(\(C_{new}\) and \(\alpha_{new}\)) with
the current value of the render target (\(C\) and \(\alpha\)). We can show that
combining batches of integration in this way
is equivalent to integrating the entire ray in a single batch.

\subsection{Rendering Styles\label{sec:visualisation:styles}}
In this particular case, we take the local Polyakov loops defined at each
lattice site and, using trilinear
interpolation, get a complex field \(L(\vec{x})\) defined everywhere on the
volume. We then calculate the complex phase
of the loops and the distance to the closest center phase:
\[\phi(\vec{x})=\arg\left(L(\vec{x})\right)\, ,\]
\[\Delta\phi(\vec{x})=\min\left(\left|\phi(\vec{x})\right|, \left|\phi(\vec{x}) 
- \frac{2\pi}{3}\right|,
\left|\phi(\vec{x}) + \frac{2\pi}{3}\right|\right)\, .\]

We then define \(C^{vol}(\vec{x})\) and \(\alpha^{vol}(\vec{x})\) to be
\[C^{vol}(\vec{x}) = \hsv\left(\frac{\phi(\vec{x})}{2\pi},1,1\right)\, ,\]
\[\alpha^{vol}(\vec{x}) =
     \begin{cases}
       100 \left(1 - 20\Delta\phi(\vec{x})\right)^4 & \text{if }
       \Delta\phi(\vec{x}) < 0.05\\
       0 & \text{if } \Delta\phi(\vec{x}) \geq 0.05
     \end{cases}
\]
where \(\hsv\) maps a color expressed in HSV (hue, saturation, value) to its
RGB representation:
\[
\hsv(h,s,v) = 
\begin{cases}
  v\cdot(1, 1-s(1-6h), 1-s) & \text{if } 0 \leq h < \frac{1}{6} \\
  v\cdot(1-s(6h-1), 1, 1-s) & \text{if } \frac{1}{6} \leq h < \frac{2}{6} \\
  v\cdot(1-s, 1, 1-s(3-6h)) & \text{if } \frac{2}{6} \leq h < \frac{3}{6} \\
  v\cdot(1-s, 1-s(6h-3), 1) & \text{if } \frac{3}{6} \leq h < \frac{4}{6} \\
  v\cdot(1-s(5-6h), 1-s, 1) & \text{if } \frac{4}{6} \leq h < \frac{5}{6} \\
  v\cdot(1, 1-s, 1-s(6h-5)) & \text{if } \frac{5}{6} \leq h < 1
\end{cases}
\]
This maps \(\phi = 0\) (\(h = 0\)) to red, \(\phi = \frac{2\pi}{3}\)
(\(h = \frac{1}{3}\)) to green,
and \(\phi = \frac{-2\pi}{3}\) (\(h = \frac{2}{3}\)) to blue.

We also use an alternative rendering style where the color is still determined
in the same way, but the
opacity is determined by the absolute value rather than the phase:
\[\alpha^{vol}(\vec{x}) =
     \begin{cases}
       200 \left(\left|L(\vec{x})\right|^2 - 0.2\right) & \text{if }
       \left|L(\vec{x})\right|^2 > 0.2\\
       0 & \text{if } \left|L(\vec{x})\right|^2 \leq 0.2
     \end{cases}
\]
This allows us to study the relationship between the phase and the absolute value.

\bibliography{reference}

\end{document}